\documentclass[a4paper,fleqn,usenatbib,twocolumn]{mnras}
\usepackage{newtxtext,newtxmath}
\usepackage[T1]{fontenc}
\usepackage{ae,aecompl}
\usepackage{graphicx}
\usepackage{amsmath}	
\usepackage{amssymb}	
\usepackage{subfig}
\usepackage{multirow}
\usepackage{tabularx}
\usepackage{array}
\usepackage{booktabs} 
\newcolumntype{P}[1]{>{\centering\arraybackslash}p{#1}}
\def\simless{\mathbin{\lower 3pt\hbox
{$\rlap{\raise 5pt\hbox{$\char'074$}}\mathchar"7218$}}}   
\def\simmore{\mathbin{\lower 3pt\hbox
{$\rlap{\raise 5pt\hbox{$\char'076$}}\mathchar"7218$}}}   
\newcommand{\be}{\begin{equation}}
\newcommand{\ee}{\end{equation}}
\title[Dust formation behind nova shocks]{Radiative shocks create environments for dust formation in classical novae}

\author[A. Derdzinski]{Andrea~M.~Derdzinski$^{1}\thanks{E-mail: aderdzinski@astro.columbia.edu}$, Brian~D.~Metzger$^{1}$, Davide Lazzati$^{2}$\\
$^{1}$Columbia Astrophysics Laboratory, Columbia University, New York, NY, 10027, USA\\
$^{2}$ Department of Physics, Oregon State University, 301 Weniger Hall, Corvallis, OR 97331, USA\\}

\begin{document}
\date{Accepted 2017 March 31}
\pagerange{\pageref{firstpage}--\pageref{lastpage}} \pubyear{2016}

\maketitle
\label{firstpage}

\begin{abstract}

Classical novae commonly show evidence of rapid dust formation within months of the outburst.  However, it is unclear how molecules and grains are able to condense within the ejecta, given the potentially harsh environment created by ionizing radiation from the white dwarf.  Motivated by the evidence for powerful radiative shocks within nova outflows, we propose that dust formation occurs within the cool, dense shell behind these shocks.  We incorporate a simple molecular chemistry network and classical nucleation theory with a model for the thermodynamic evolution of the post-shock gas, in order to demonstrate the formation of both carbon and forsterite ($\rm Mg_2SiO_4$) grains.  The high densities due to radiative shock compression ($n \sim 10^{14}$ cm$^{-3}$) result in CO saturation and rapid dust nucleation.  Grains grow efficiently to large sizes $\gtrsim 0.1\mu$m, in agreement with IR observations of dust-producing novae, and with total dust masses sufficient to explain massive extinction events such as V705 Cas.  As in dense stellar winds, dust formation is CO-regulated, with carbon-rich flows producing carbon-rich grains and oxygen-rich flows primarily forming silicates.  CO is destroyed by non-thermal particles accelerated at the shock, allowing additional grain formation at late times, but the efficiency of this process appears to be low.  Given observations showing that individual novae produce both carbonaceous and silicate grains, we concur with previous works attributing this bimodality to chemical heterogeneity of the ejecta.  Nova outflows are diverse and inhomogeneous, and the observed variety of dust formation events can be reconciled by different abundances, the range of shock properties, and the observer viewing angle.  The latter may govern the magnitude of extinction, with the deepest extinction events occurring for observers within the binary equatorial plane.

\end{abstract} 
  
\begin{keywords}
novae, cataclysmic variables -
shock waves -
\end{keywords}

\section{Introduction} 
\label{sec:introduction}

Classical and recurrent novae are explosive events, powered by a thermonuclear runaway (TNR) on the surface of an accreting white dwarf (WD) in a semi-detached binary. They reach peak luminosities close to or exceeding the Eddington luminosity, and they eject a total mass of $\sim 10^{-5}-10^{-4}M_{\odot}$ at velocities of hundreds to thousands of km s$^{-1}$ (e.g., \citealt{Bode&Evans08}; \citealt{Gehrz08}).  Although a fraction of the matter is unbound promptly following the TNR, the remainder emerges in a continuous wind from the WD (e.g.~\citealt{Kato&Hachisu94}).  Novae remain bright optical sources for weeks to months, but even after they fade their bolometric luminosities can remain roughly constant for years to decades (\citealt{Krautter08}).  As the pseudo-photosphere recedes back through the ejecta shell, the emission temperature rises, shifting to ultraviolet and then soft X-ray frequencies.  

Novae are remarkably diverse in their composition, spectral development, and the decline rate of their visual light curves  (see \citealt{Shore12} for a review). Starting with DQ Her in 1934 (\citealt{McLaughlin35}), some nova light curves show dramatic minima, indicating the sudden formation of dust within the ejecta.  These events are typically observed $20-100$ days after the outburst, accompanied by a simultaneous increase in the mid-infrared emission.  This picture was confirmed in the 1970s with broadband infrared (IR) photometry (\citealt{Geisel+70}; \citealt{Ney&Hatfield78}; \citealt{Gehrz+80}), which shows that the IR light curve usually rises to a maximum over a few months after the outburst (see \citealt{Evans&Gehrz12} for a review). 
Dust grains in novae often grow to large sizes (e.g. as inferred by SED modeling of V1065 Cen, where grain sizes reach up to $\sim 1 \mu \rm m$). 

While only $\sim \! 20\%$ of classical novae form dust along the line of site, at least $40\%$ show signatures of dust in the IR, indicating that dust forms with a partial covering fraction (\citealt{Helton2010}; F.~Walter, private communication).  Dust formation is more common in novae from less massive CO white dwarfs, and is nearly absent in the Ne-enriched novae which are thought to originate from more massive ONe WDs (\citealt{Evans&Gehrz12}), with some notable exceptions, e.g. QU Vul (\citealt{Schwarz02}) and V1065 Cen \citep{Helton10}.  When dust does form, its mass fraction within the ejecta is typically $X_{\rm d} \sim 10^{-3}$, corresponding to a total dust mass of $\sim 10^{-7}-10^{-6}M_{\odot}$ as inferred, e.g., from the nova remnant GK Per (\citealt{Bode+87}, \citealt{Dougherty+96}).  

Optical and ultraviolet studies, as well as theoretical modeling of the TNR, indicate that the ejecta of classical nova is typically rich in oxygen (\citealt{Gehrz+98}; \citealt{Starrfield+98}), perhaps predicting a preference for the formation of silicate dust.  However, the same novae which show silicate dust features, in particular the well known 9.7$\mu$m SiO stretch feature (e.g.~\citealt{Gehrz+85}, \citealt{Roche+84,Smith+95}), often show evidence for carbon grains.  In particular, some novae show evidence for emission features which are potentially associated with amorphous carbon (\citealt{Snijders+87}) or hydrogenated amorphous carbon (\citealt{Scott+94}).   

The bimodality of carbonaceous and silicate dust in novae may be intimately tied to the chemistry of carbon monoxide (CO). When early near-infrared spectroscopic observations are available, the molecule is often detected within days of the optical outburst and prior to the formation of dust (e.g.~\citealt{Evans+96}; \citealt{Rudy+03}).  According to the conventional paradigm based on dense stellar winds, the composition of dust that forms in a given environment depends sensitively on the ratio of carbon to oxygen.  When the C:O ratio exceeds unity (carbon-rich), carbonaceous dust forms, while silicate dust forms when C:O ratio is less than unity (e.g.~\citealt{Waters04}).  This expectation is based on the assumption that the CO abundance reaches its saturation value, which may not always be a good approximation for the conditions in the nova ejecta.  Carbon nucleation can occur even in an oxygen rich environment if free carbon is made available by CO destruction, as can occur due to neutral reactions in a shielded region (\citealt{Pontefract&Rawlings04}) or due to dissociation by energetic electrons (e.g. \citealt{Todini&Ferrara01,Clayton2013,Lazzati&Heger16}).  In core collapse supernovae, energetic electrons are produced by the radioactive decay of $^{56}$Ni.  As we shall describe, in novae the ionizing particles are instead more likely to originate from shocks.

At first glance, the hostile environment of a nova is not an ideal one to form molecules and dust.  The chemistry leading to the formation of first diatomic, then polyatomic molecules and dust, requires an environment that is shielded from the hard radiation of the white dwarf (\citealt{Bath&Harkness89,Johnson93}).  The latter remains a supersoft X-ray source radiating near the Eddington luminosity for years or longer after the eruption (e.g.,~\citealt{Schwarz+11}).  The general consensus appears to be that dust formation necessitates a carbon neutral region  (\citealt{Rawlings&Williams89}; \citealt{Rawlings88}, \citealt{Rawlings&Evans08}).  However, this is only possible in regions of the ejecta where the gas density is much higher than the average value one would predict based on free homologous expansion.  Such density enhancements are supported empirically by spectroscopic modeling of novae which show evidence for a radially thin shell of clumpy ejecta (e.g.~\citealt{Williams92}, \citealt{Saizar&Ferland94}), but their physical origin has not (to our knowledge) been previously addressed (except in the rare case of recurrent novae that collide with remnants of previous eruptions, such as seen in T Pyxidis, e.g. \citealt{Toraskar13}).

In the traditional view, novae are powered directly by the energy released from nuclear burning  (e.g.~\citealt{Hillman+14}).  However, growing evidence suggests that shock interaction plays an important role in powering nova emission across the electromagnetic spectrum.  This evidence for shocks includes multiple velocity components in the optical spectra (e.g.~\citealt{Williams&Mason10}); hard X-ray emission starting weeks to years after the outburst (e.g.~\citealt{Mukai+08}); and an early sharp maximum in the radio light curve on timescales of months, in clear excess of that expected from freely-expanding photo-ionized ejecta (e.g.~\citealt{Chomiuk+14}; \citealt{Weston+15}).  

The most striking indicator of shocks in novae is the recent discovery by {\it Fermi} LAT of $\gtrsim$ 100 MeV gamma-rays, observed at times coincident within a few days of the optical peak and lasting a few weeks (\citealt{Ackermann+14}, \citealt{Cheung+2016}). Continuous gamma ray emission indicates the presence of strong shocks that are able to accelerate particles (via diffusive shock acceleration) to relativistic energies by leptonic and hadronic processes. In the leptonic case, electrons  accelerated to relativistic energies by the shock can then inverse Compton scatter surrounding photons to GeV energies. In the hadronic case, accelerated protons will collide with ambient protons in the post-shock gas, producing high-energy electron-positron pairs as well as neutral pions that decay to gamma rays.  (See \citealt{Metzger2015} for a detailed study of non-thermal emission from nova shocks.)

The first nova with detected gamma-rays occurred in the symbiotic binary V407 Cyg \citep{Abdo+10}, suggesting that shocks were produced by the interaction between the nova outflow and the dense wind of the companion red giant.  However, gamma-rays have now been detected from more than five {\it ordinary} classical novae with main sequence companions (\citealt{Ackermann+14}, \citealt{Cheung+2016}).  Remarkably, this demonstrates that the nova outflow runs into dense gas even in systems not embedded in the wind of an M giant or associated with recurrent novae.  This dense gas instead likely represents lower velocity mass ejected earlier in the outburst (`internal shocks'; \citealt{Friedjung87}; \citealt{Mukai&Ishida01}, \citealt{Metzger+14}).  Current observations are consistent with many, and possibly all, novae producing shocks and $\gtrsim$ 100 MeV gamma-ray emission.  The LAT-detected novae appear to be distinguished primarily by their relatively close distances (e.g., \citealt{Finzell+15}).  
 
Due to the high densities of novae ejecta, shocks in this environment are likely to be radiative (\citealt{Metzger+14}, \citealt{Metzger+15}).  As gas cools behind the radiative shock, its density increases by a factor of $\gtrsim 10^{3}$.  Here we show that this provides an ideal environment for forming dust, while simultaneously offering a natural explanation for the extreme density inhomogeneities observed within nova ejecta.  

This paper is organized as follows.  In $\S\ref{sec:shocks}$ we discuss the dynamics of radiative shocks and describe our model for the thermodynamic trajectories experienced by the post-shock gas.  In $\S\ref{sec:dust}$ we describe a model for molecule formation and dust nucleation, and in $\S\ref{sec:results}$ we present the results, considering variations from our fiducial model. We discuss the implications of our results in $\S\ref{sec:discussion}$, with comparisons to observations.  For a brief summary of our conclusions, see $\S\ref{sec:conclusions}$.    

\section{Radiative Shocks Create Dust Formation Sites}

\label{sec:shocks}

\subsection{Shock Geometry and Dynamics}

\begin{figure}
\includegraphics[width=0.5\textwidth]{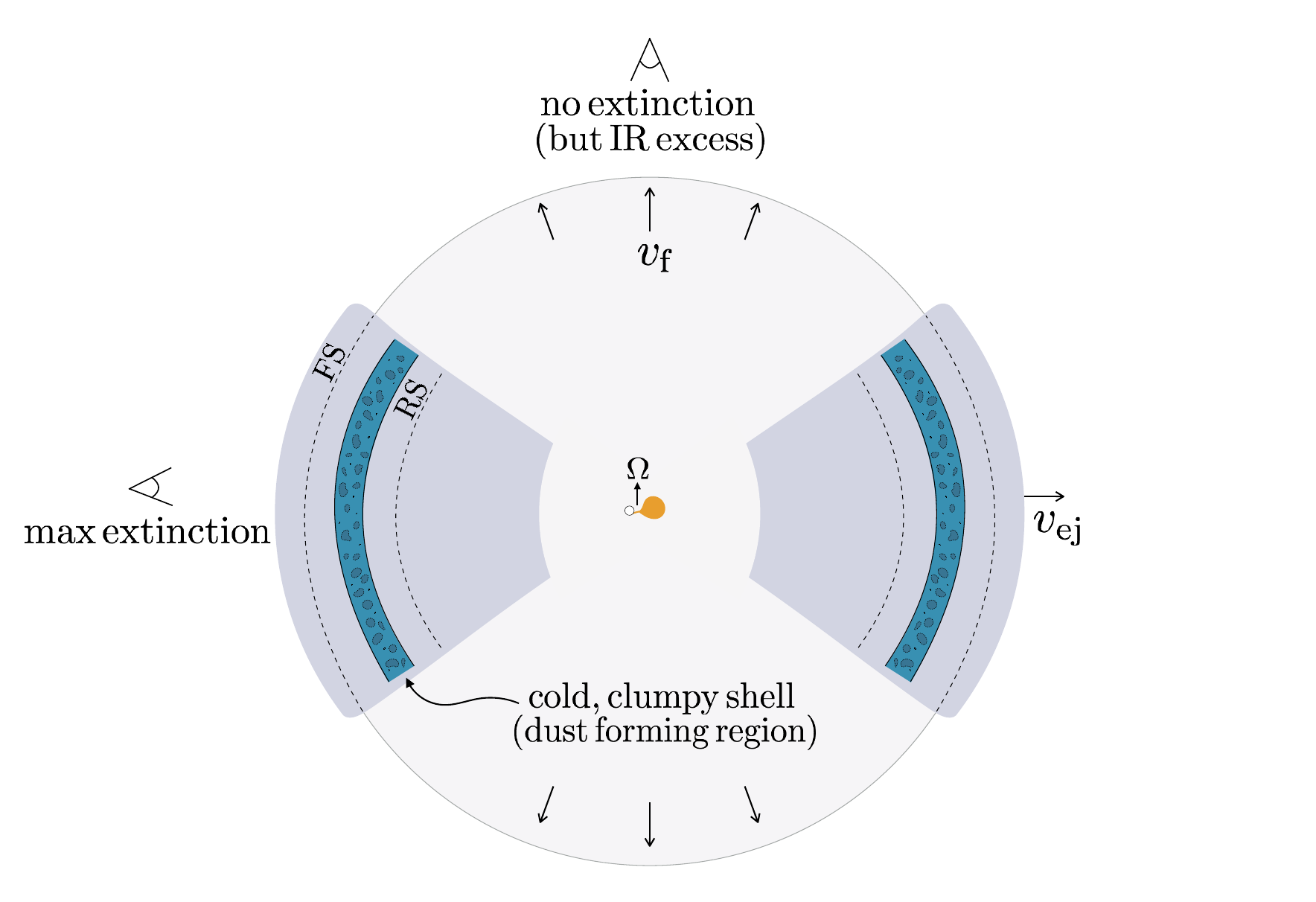}
\caption{Schematic diagram of the nova ejecta. A dense, equatorially-concentrated outflow is followed by a faster, isotropic wind that collides with the slow ejecta, producing shocks in the equatorial plane. The post-shock gas is able to cool efficiently, producing a cool, dense, (likely clumpy) shell between the forward and reverse shocks. This neutral region is precisely where we expect dust to form. In this geometry, the deepest extinction events will occur for observers within the equatorial plane of the binary. }
\label{fig:diagram}
\end{figure}

Shocks were unexpected in classical novae because the pre-eruption environment surrounding the white dwarf is occupied only by the low density wind of the main sequence companion star, requiring a different source of matter into which the nova outflow collides.  One physical picture, consistent with both optical (e.g., \citealt{Schaefer+14}) and radio imaging (e.g., \citealt{Chomiuk+14}), and the evolution of optical spectral lines (e.g.,~\citealt{Ribeiro+13}; \citealt{Shore+13}), is that the thermonuclear runaway is first accompanied by a slow ejection of mass focused in the equatorial plane, the shape of which may be influenced by the binary companion (e.g., \citealt{Livio+90}; \citealt{Lloyd+97}).  This slow outflow is then followed by a second ejection or a continuous wind (e.g., \citealt{Bath&Shaviv76}) with a higher velocity and more spherical geometry.  The subsequent collision between the fast and slow components produces strong ``internal" shocks within the ejecta which are concentrated in the equatorial plane.  The fast component continues to expand freely along the polar direction, creating a bipolar morphology. Figure~\ref{fig:diagram} shows a schematic diagram of the proposed geometry, with the cold, clumpy dust-forming region highlighted in blue (see also \citealt{Sokoloski+08}, \citealt{Orlando+09}, \citealt{Drake+09}, \citealt{Orlando&Drake12} for other evidence for bipolar ejecta in symbiotic novae).

We begin by estimating the density and radius of the shocked gas on the timescale of weeks - a month of the observed gamma-rays. The slow outflow velocity is typically $v_{\rm ej} = 1000$ km s$^{-1}$, which we show with the notation $v_{\rm ej} = 10^{3}v_{\rm ej,8}$ km s$^{-1}$ where $v_{\rm ej,8} = v_{\rm} / 10^8$ cm s$^{-1}$. This initial ejecta expands to a radius $R_{\rm ej} = v_{\rm ej} t \approx 6\times 10^{13}t_{\rm wk}v_{\rm ej,8}\,{\rm cm}$ by the time $t_{\rm wk} =t / (6\times10^5) $ s.  
The characteristic density of the unshocked ejecta of assumed thickness $\sim R_{\rm ej}$ can be estimated as
\be
n_{\rm ej} \approx \frac{M_{\rm ej}}{4\pi R_{\rm ej}^{3}f_{\Omega} m_p} \sim 9\times 10^{10}M_{-4}t_{\rm wk}^{-3}v_{\rm ej, 8}^{-3}\,{\rm cm^{-3}},
\label{eq:nej}
\ee
where $f_{\Omega} \sim 0.5$ is the fraction of the total solid-angle subtended by the outflow and $M_{\rm ej} = 10^{-4}M_{-4}M_{\odot}$ is the ejecta mass, normalized with a characteristic value  $M_{-4} = M_{\rm ej} / 10^{-4} M_{\odot}$ (e.g., \citealt{Seaquist+80}).  

A faster outflow (``wind") of mass loss rate $\dot{M}$ and velocity $v_{\rm f} > v_{\rm ej}$ collides with the ejecta from behind.  The density of the wind at the collision radius ($\sim$ radius of the slow ejecta) is given by
\be
n_{\rm w} \approx \frac{\dot{M}}{4\pi R_{\rm ej}^{2}m_p v_{\rm f}} \sim 2\times 10^{9}\dot{M}_{-4}v_{\rm ej,8}^{-3}(v_{\rm f}/v_{\rm ej})^{-1}t_{\rm wk}^{-2}{\,\rm cm^{-3}},
\label{eq:nw}
\ee
where the mass loss rate $\dot{M} = 10^{-4}\dot{M}_{-4}M_{\odot}\,{\rm wk^{-1}}$ is normalized to a value resulting in the ejection of $\sim 10^{-4}M_{\odot}$ over a week.  

This interaction drives a forward shock through the slow ejecta and a reverse shock back through the fast ejecta, with a shock velocity $v_{\rm sh}$ we normalize by $v_{\rm 8} = v_{\rm sh}/10^{8}$ cm s$^{-1}$.
Shocks heat the gas to a temperature
\be
T_{\rm sh} \simeq \frac{3}{16 k}\mu m_p v_{\rm sh}^{2} \approx 1.7\times 10^{7}v_{\rm  8}^{2}\,{\rm K},
\label{eq:Tsh}
\ee
and compresses it to a density $n_{\rm sh} = 4n$, where $n = n_{\rm ej}$ (forward shock) or $n = n_{\rm w}$ (reverse shock).  Here $\mu \simeq 0.76$ is the mean molecular weight assuming a fully ionized gas with a composition of our fiducial model. 

If the shocks are radiative, the conditions for which are described below, then the post shock material cools rapidly and piles up in a central cold shell\footnote{In fact, the central shell is likely to be highly inhomogeneous due to well-known instabilities associated with radiative shocks (e.g.~\citealt{Chevalier&Imamura82,Vishniac83}).  Such inhomogeneities may explain the `clumpiness' inferred from spectroscopic studies of nova ejecta, but we neglect this complication for now. } sandwiched by the ram pressure of the two shocks (\citealt{Metzger+14}).  The shocked gas cools by a factor of $\sim 10^{3}$, its volume becoming negligible ($\S\ref{sec:thermo}$).  The post-shock gas cools so fast that it supplies little pressure support; for this reason, the shocks propagate outwards at the same velocity as the central shell, $v_{\rm sh}$, which obeys $v_{\rm ej} \lesssim v_{\rm sh} \lesssim v_{\rm f}$.  The existence of a cold thin shell composed of many `clumps' or condensations is an observed feature of the ejecta structure of classical novae (e.g.~\citealt{Saizar&Ferland94,Friedjung+99}).  

The cooling timescale of the post-shock gas is given by
\be
t_{\rm cool} = \frac{3kT_{\rm sh}}{2\mu \Lambda n_{\rm sh}},
\label{eq:tcool}
\ee 
 where the cooling function comes from combining our fiducial abundances with the radiative cooling curves presented in \citealt{Schure2009}.  Following \citealt{Vlasov+2016}, the cooling function  $\Lambda(T) = \Lambda_{\rm lines} + \Lambda_{\rm ff}$ contains contributions from free-free emission  
\be
 \Lambda_{\rm ff} \approx 1\times 10^{-23}(T/10^{7}{\rm K})^{1/2}\,\,\, {\rm erg\,\, cm}^{3}\,\, {\rm s}^{-1} \label{eq:lambdaff}
\ee
and from emission lines, $\Lambda_{\rm lines}$. The latter can be approximated in the temperature range $3\times 10^{5}K \lesssim T \lesssim 10^{7}K$ as (\citealt{Vlasov+2016}).
\be
 \Lambda_{\rm lines} \approx
  2\times 10^{-22}\left(\frac{T}{10^7{\rm K}}\right)^{-0.7}{\rm erg\,cm^{3}\,s^{-1}},
 \label{eq:lambda} 
\ee
where the normalization depends on the gas composition and is generally much higher than for solar metallicity gas due to the large enhancements of CNO elements.  Line cooling dominates free-free cooling for shock temperatures $T \lesssim 3\times 10^{7}$ K, corresponding to shock velocities $v_{\rm sh,8} \lesssim 1$.
 
Whether the shocks are radiative depends on the ratio of the cooling timescale to the expansion timescale $\sim t$.  Focusing on the forward shock, we find that
\begin{eqnarray}
&& \frac{t_{\rm cool}}{t} \approx
\left\{
\begin{array}{lr}
 3.1\times 10^{-3}v_{\rm 8}^{4}M_{-4}^{-1}t_{\rm wk}^{2}
 &
v_8 \lesssim 1, \\
3\times 10^{-4}  v_{\rm 8}^{5.4}M_{-4}^{-1}t_{\rm wk}^{2}&
v_8 \gtrsim 1, \\
\end{array}
\right.
\label{eq:cool}
\end{eqnarray}
where we have used eq.~(\ref{eq:tcool}) along with eqs.~(\ref{eq:lambdaff}), (\ref{eq:lambda}) in the bottom and top limits, respectively.

For shock velocities $v_{\rm 8} \lesssim 1$ and ejecta masses $M_{-4} \sim 0.1-1$, the forward shock is likely to be radiative, at least during the epoch of highest gamma-ray luminosity.  Gas cools behind the shock on the characteristic cooling length scale $L_{\rm cool} = (v_{\rm sh}/4)t_{\rm cool}$, such that the radiative shock condition can also be written as $L_{\rm cool} \ll R_{\rm ej}/4$.

In summary, the nova shocks responsible for the gamma-ray emission are characterized by upstream (pre-shock) gas densities $n \sim 10^{9}-10^{11}$ cm $^{-3}$ and velocities $v_{\rm sh} \sim 500-2000$ km s$^{-1}$.  However, these ranges remain uncertain because the outflow and shock geometry is poorly understood theoretically and is observationally unresolved at the time of the gamma-ray emission.  However, it is important to note that the shock properties are constrained to produce the high observed gamma-ray luminosities \citep{Ackermann+14} of
\be
L_{\gamma} = \epsilon_{\rm nth}\epsilon_{\gamma}L_{\rm sh} \sim 10^{35}-10^{36} {\rm erg\,\, s^{-1}},
\label{eq:Lgamma}
\ee
where 
\be L_{\rm sh} = (9\pi/8)f_{\Omega} R_{\rm ej}^{2}n m_p v_{\rm sh}^{3} \gtrsim 10^{37}-10^{38}\,{\rm erg\,s^{-1}}
\label{eq:Lsh}
\ee 
is the kinetic power dissipated by the shocks, $n$ is the density of the upstream unshocked gas ($n_{\rm ej}$ or $n_w$ above), $\epsilon_{\rm nth} \lesssim 0.1$ is the fraction of the shock power used to accelerate relativistic non-thermal particles, and $\epsilon_{\gamma} \lesssim 0.1$ the fraction of this energy radiated in the LAT bandpass (\citealt{Metzger+15}).  For typical values of the shock radius and density, one finds that {\it an order unity fraction of the ejecta must pass through a shock in order to explain the high observed gamma-ray luminosities.}  The has important implications for the fraction of observers likely to detect dust extinction events ($\S\ref{sec:discussion}$).

\subsection{Thermodynamic Evolution of Post Shock Gas}
\label{sec:thermo}

The pressure behind the shock is given by the jump conditions, 
\be
P_{\rm sh} = \frac{2L_{\rm sh} }{3\pi v_{\rm sh} R_{\rm ej}^{2}}.
\label{eq:Psh}
\ee
The post-shock gas forms a cold shell at a temperature $T_{\rm cs}$. For an ideal gas\footnote{We neglect non-thermal pressure support in the post-shock region due to the magnetic field or relativistic particles, which can act to temporarily halt the compression of the post-shock gas.  However, the (small-scale, shock-generated) magnetic field is likely to decay away downstream of the shock.  Non-thermal particles will also eventually cool via Coulomb collisions and high energy radiation (inverse Compton, synchrotron, and relativistic bremsstrahlung emission), after which time the compression will proceed as if it were controlled exclusively by thermal cooling \citep{VurmMetzger2016}.} 
which cools to a temperature $T_{\rm cs} =  10^{4}T_{\rm cs,4}$ K (see eq.~\ref{eq:Tsh}) at roughly constant pressure $P = nkT$, the density increases by a factor of $\beta \equiv 4 T_{\rm sh}/T_{\rm cs} \approx 9\times 10^{3}v_{8}^{2}T_{\rm cs,4}^{-1}$ from its pre-shock value.  Assuming that a significant fraction of the ejecta mass is shocked, this matter collects into a  cold shell of mass $M = M_{-4}10^{-4} M_{\odot}$ of thickness $\Delta \approx R_{\rm ej}/\beta$ and characteristic density
\be
n_{\rm max} \approx \frac{M}{4\pi R_{\rm ej}^{2}\Delta m_p} \underset{R_{\rm sh} \sim t v_{\rm sh}}\approx 4\times 10^{14}t_{\rm wk}^{-3}v_{8}^{-1}M_{-4}T_{\rm cs,4}\,{\rm cm^{-3}},
\ee   
For typical values of $M_{-4} \sim 0.1-1$ and $t_{\rm wk} \sim 1-2$, the density reaches a maximum value of $n_{\rm max} \sim 10^{13}-10^{14}$ cm$^{-3}$.  After achieving this maximum density, cooling slows due to radiative heating by the nova light (see below).  After this point, both lateral expansion (due to the increasing radius of the shell) and radial expansion (as the shock pressure subsides) will cause the density of the shell to decrease from its maximum value of $n_{\rm max}$.

We now describe a more detailed model for the thermodynamic evolution of a fluid element which enters the shock and then is deposited in the central shell.  Given its high inertia, the central shell is assumed to expand ballistically, with $R_{\rm sh} = v t \simeq R_{\rm ej}$ and velocity $v$ (\citealt{Metzger+14}), which for concreteness we take to be twice the shock velocity $v_{\rm sh}$.  The kinetic luminosity of the shock is assumed to evolve with time as 
\be
L_{\rm sh} = \frac{L_{\rm sh,0}}{1 + (t/\tau_0)^{\alpha}},
\label{eq:Lsh2}
\ee
where $\tau_0 = 1.5\times 10^{6}$ s is a characteristic decay timescale chosen to approximate the duration of the observed the gamma-ray emission, and the peak luminosity $L_{\rm sh,0} = 10^{38}$ erg s$^{-1}$ is that required to power the observed gamma-ray luminosities (\citealt{Ackermann+14}; eq.~\ref{eq:Lsh} and surrounding discussion).  Although the precise form of the temporal decay of $L_{\rm sh}$ we have adopted is somewhat ad hoc, our choice of $\alpha = 4$ results in the shock luminosity being approximately equal to that needed to power the non-thermal radio emission on a timescale of a few months (\citealt{Metzger+14}; \citealt{Vlasov+2016}).\footnote{Radio emission, initially absorbed by photo-ionized gas ahead of the shock, only peaks once the density of the shocked gas has decreased by a factor of $\sim 10^{3}$ from its peak value at the time of the gamma-ray emission.}  

The enthalpy of the shell evolves with time according to
\be
\frac{d}{dt}\left(\mathcal{E} + \frac{P_{\rm sh}}{\rho}\right) = \frac{dP_{\rm sh}}{dt}\frac{1}{\rho} - \frac{L_{\rm rad}}{M},
\label{eq:enth}
\ee
where the shock pressure $P_{\rm sh}$ follows from the shock luminosity from equation (\ref{eq:Psh}), $\mathcal{E}$ is the specific internal energy, $\rho = m_p n$ is the density, and
\be
L_{\rm rad} = \frac{16\pi \sigma R_{\rm sh}^{2}T^{4}}{9(\tau + 2/(3\tau) + 4/3)}
\label{eq:Lrad}
\ee
is the radiative luminosity of the shell (e.g.~\citealt{Sirko&Goodman03}),
\be
\tau = \frac{M\kappa}{8\pi R_{\rm sh}^{2}},
\ee 
is the optical depth from the center of the shell, where $\kappa(\rho,T)$ is the Rosseland mean opacity.  Our approximate expression for $L_{\rm rad}$ is borrowed from the accretion disk literature and smoothly interpolates between the optically-thin ($\tau \ll 1$) and optically-thick regimes ($\tau \gg 1$).  In the optically-thin regime, this expression reduces to the volume of the shell $4\pi R_{\rm sh}^{2}\Delta$ (where $\Delta$ is the shell thickness) times the frequency-integrated volume emissivity $\int j_{\nu} d\nu$, where in LTE $j_{\nu}/\kappa_{\nu} = B_{\nu}(T)$, $B_{\nu}(T)$ is the blackbody source function, and $\kappa_{\nu}$ is the absorption coefficient (such that $\tau \approx \Delta \int \kappa_{\nu}d\nu$).  In the optically-thick regime, equation \ref{eq:Lrad} reduces to the solution of the diffusion equation for radiation out of a 1D slab of optical depth $\tau$ and central temperature $T$.

We solve equation (\ref{eq:enth}) given the known evolution of $dP_{\rm sh}/dt$, starting from a time $t_{\rm shock}$, corresponding to the time after the outburst when a `typical' fluid element enters the shock.  The initial conditions at $t = t_{\rm shock}$ are specified by the initial temperature $T(t_{\rm shock}) \gtrsim 10^{4}$ K, which given the initial pressure determines the initial enthalpy.  Because the cooling time behind the shock is originally so short, our results are not sensitive to the precise value of $T(t_{\rm shock})$ as long as it exceeds the temperature range of several thousand K relevant to molecule and dust formation.

At each time step, the known pressure $P_{\rm sh}$ and enthalpy fully determine the other thermodynamic variables (temperature and density) given an equation of state (EOS).  We adopt the EOS of  \citet{Tomida+13,Tomida+15}, which takes into account ionization of hydrogen and helium and molecular states of $H_2$.  We assume a hydrogen composition of $X_{\rm H} = 0.4$ with the remainder in helium, motivated by our fiducial abundances ($\S\ref{sec:abundances}$) and adopt a ratio of ortho- to para-$H_2$ equal to its equilibrium value.  We employ Rosseland mean opacities as compiled by \citet{Tomida+13}, which are based on \citet{Semenov+03}, \citet{Ferguson+05} and the Opacity Project (\citealt{Seaton+94}).

Equation (\ref{eq:enth}) neglects radiative {\it heating} due to the nova luminosity streaming through the central shell.  Following \citet{Pontefract&Rawlings04}, we account for this by setting a floor on the temperature of
\be
T_{\rm rad} \simeq \left(\frac{L }{4\pi R_{\rm sh}^{2}\sigma}\right)^{1/4}\!\! \underset{R_{\rm sh} \sim v_{\rm sh}t}\approx 2500\,{\rm K}\,\left(\frac{L}{10^{38}\,\rm erg\,s^{-1}}\right)^{1/4} \! v_{8}^{-1/2}t_{\rm wk}^{-1/2}
\label{eq:Trad}
\ee
where $L = 10^{38}$ erg s$^{-1}$ is the WD luminosity, which we assume is temporally constant throughout the dust formation epoch.  

\begin{figure}
\subfloat{
\includegraphics[width=0.5\textwidth]{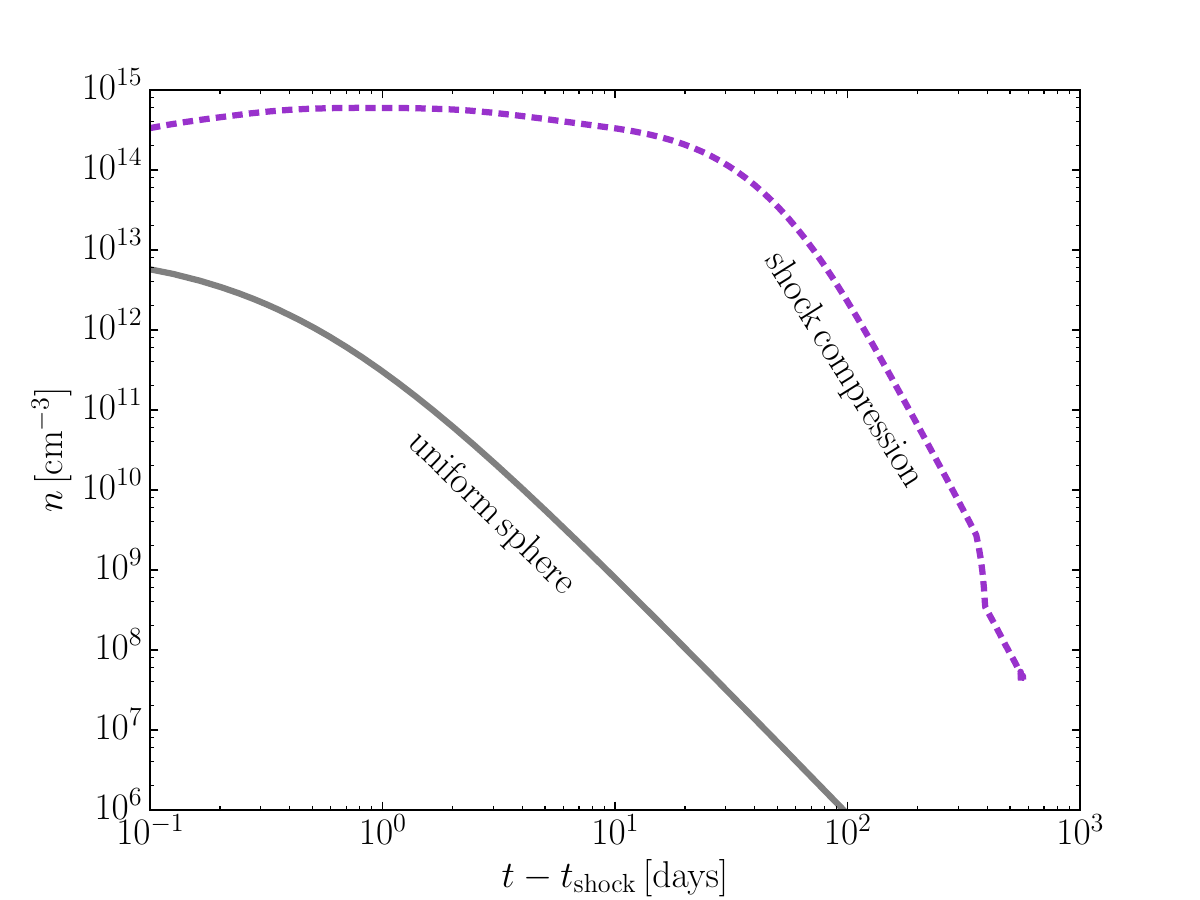}}

\vspace{-4mm}

\subfloat{
\includegraphics[width=0.5\textwidth]{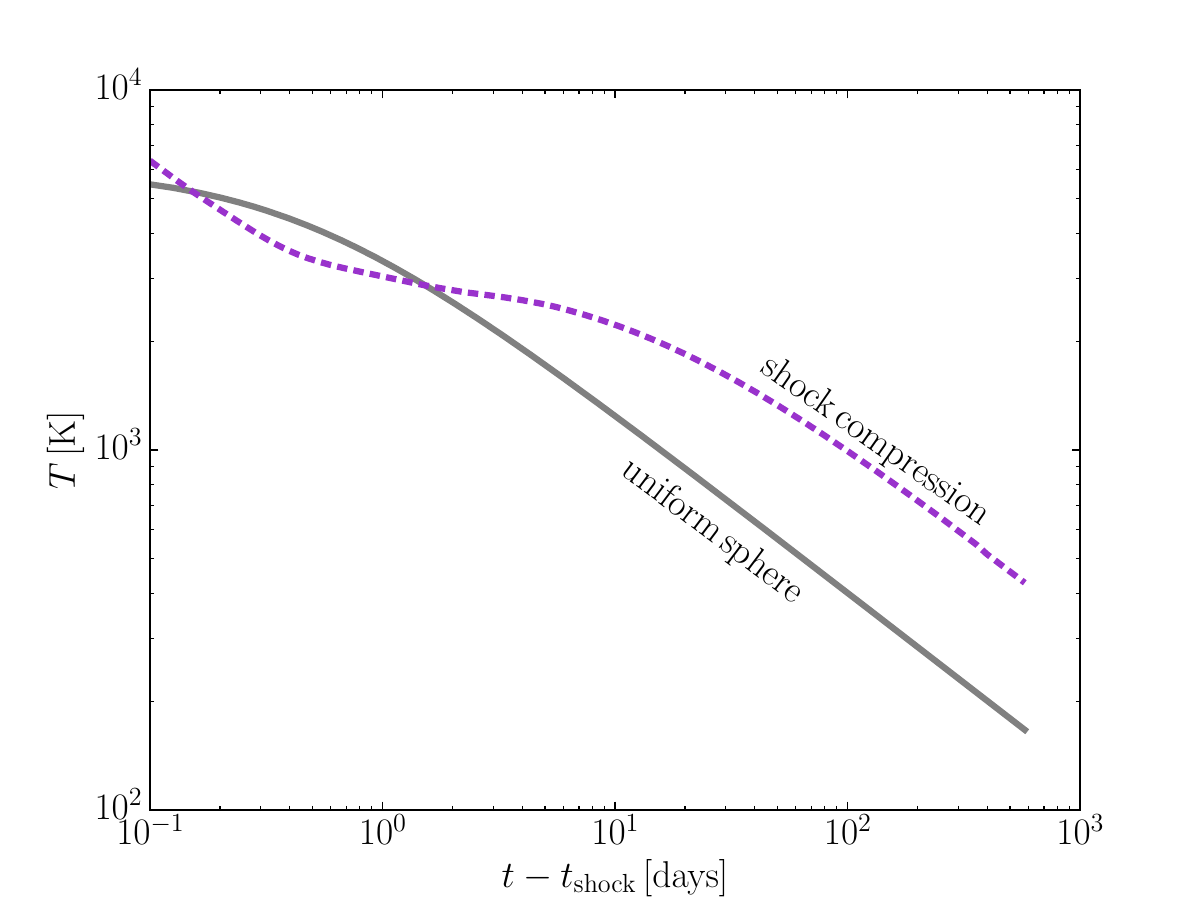}}

\vspace{-4mm}

\subfloat{
\includegraphics[width=0.5\textwidth]{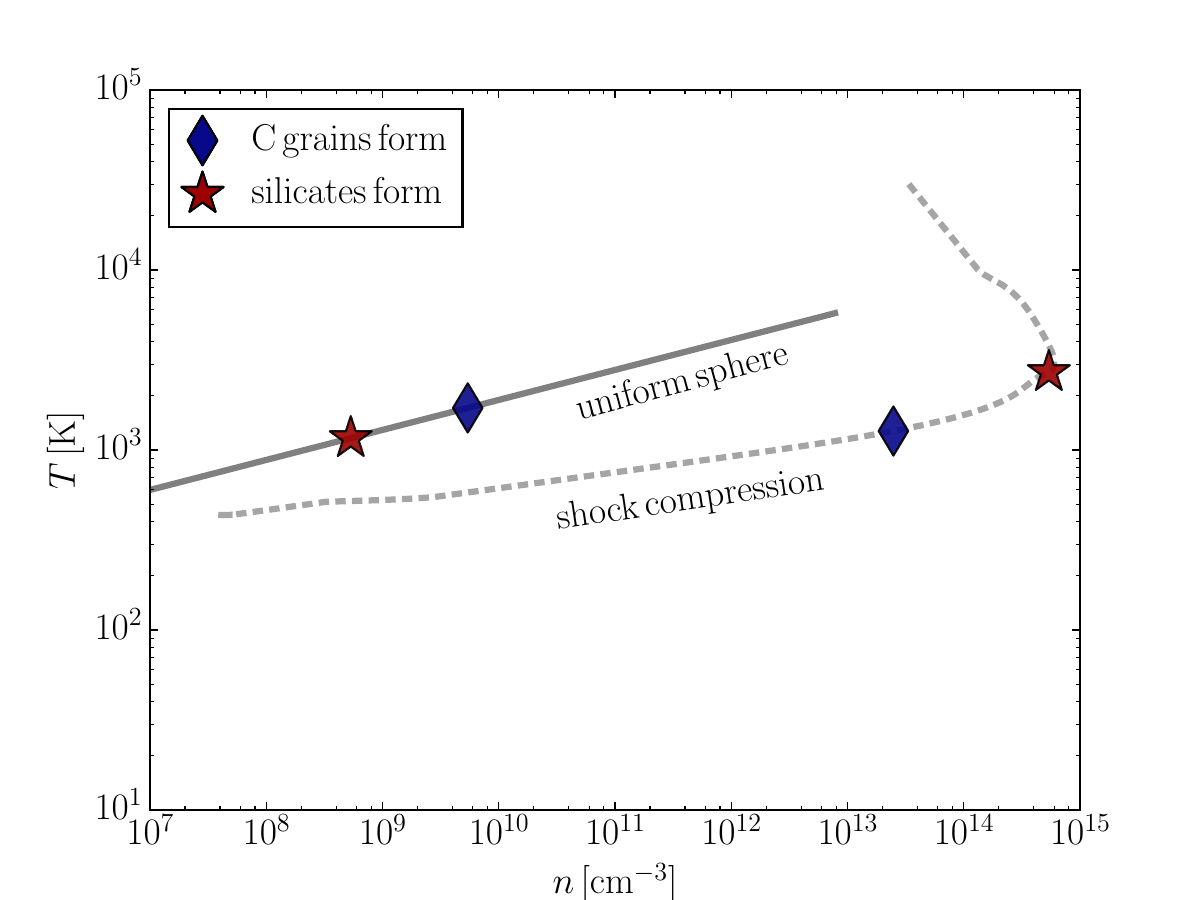}}
\caption{Density and temperature evolution for the post-shock gas, calculated for mean ejecta velocity $v_{\rm ej} = 500$ km s$^{-1}$ and shock velocity $v_{\rm sh} = 500$ km s$^{-1}$ .  The fluid element enters the shock at a time $t_{\rm shock}\sim10^{6}$ s.  Our fiducial trajectory (`shock compression', dashed line) includes radiative cooling and heating from the WD. For comparison, we include the evolution of a freely expanding uniform sphere that is irradiated by the WD luminosity ('uniform sphere', grey line).
We also show the thermodynamic evolution in the temperature-density plane, where symbols indicate at what density and temperature carbon grains (blue diamonds) and silicate grains (red stars) condense. A notable feature in the trajectory involving compression by shocks is the formation of grains at very high densities, where the gas is necessarily neutral.  }
\label{fig:thermo}
\end{figure}

Figure \ref{fig:thermo} shows the thermodynamic trajectories for our fiducial model of $T(t)$, $n(t)$, and $T(n)$, calculated for an ejecta velocity of $v_{\rm ej} = 500$ km s$^{-1}$, shell velocity $v = 1000$ km s$^{-1}$ and resulting shock velocity of $v_{\rm sh} \simeq v - v_{\rm ej} = 500$ km s$^{-1}$.  At early times, gas cools rapidly from temperatures $T \gtrsim 10^{4}$ K to $T \sim 2-3000$ K as the density concomitantly rises.  Below this temperature, radiative cooling slows down and adiabatic expansion instead begins to dominate the loss of internal energy.  After this point catastrophic cooling is prevented because the temperature has reached its floor set by external radiation.

Absent shocks, the bulk of the ejecta, which is released on a timescales of days to weeks after the outburst, would be expected to approach a phase of free radial homologous expansion on a timescale of months when dust formation is observed.  The simplest assumption would be to approximate its geometry as that of a freely expanding uniform sphere, with the density decreasing as $n = {M_{\rm ej}}/{4 \pi R_{\rm ej}^3 m_p} \propto t^{-3},$ and the temperature determined by irradiation from a central source according to equation (\ref{eq:Trad}).  For comparison to our shock models, we also consider trajectories corresponding to such a expanding uniform sphere, as shown with a solid grey line in Figure~\ref{fig:thermo}.

A key question relevant to dust formation is whether the ejecta will remain neutral as it expands, or whether it will become ionized by X-rays from the central white dwarf.  In Appendix \ref{sec:ionization}, we estimate that carbon will remain neutral until the density decreases below a critical value of $n_{\rm ion} \approx 10^{9}-10^{10}$ cm$^{-3}$, depending on the WD luminosity, the carbon abundance, and the mass of the central shell.  Figure \ref{fig:thermo} shows that for characteristic parameters these densities are not reached in our shock models until very low temperatures $T \lesssim 500$ K, well after dust condensation and growth.  By comparison, for thermodynamic conditions which correspond to the naive picture of a uniform homologous sphere (i.e. neglecting shocks), these densities are reached when the temperature is much higher, $T \gtrsim 2000$ K, i.e.~before dust would have formed.  Given the likely detrimental effects of ionizing radiation on the dust formation process, this already illustrates the crucial role played by enhanced compression due to radiative shocks in creating an environment conducive to dust formation.

\section{Dust Formation Model}
\label{sec:dust}

In this section we describe a model for the formation of simple molecules and concomitant dust nucleation.

\subsection{CO and SiO Chemistry }
\label{sec:chemical}
Molecule formation is not only a necessary precursor to dust formation; it also plays an important role in regulating abundances. While molecules must exist prior to forming grains, they can also prevent nucleation by locking up the needed elements. For these reasons, we include the non-equilibrium formation and destruction of both SiO and CO. CO is particularly inert within the temperature range of interest, serving as an efficient sink for free carbon before nucleation can take place, while SiO is a precursor molecule for silicate dust (Section~\ref{sec:dust}). 

Without assuming chemical equilibrium, the abundances of CO and SiO evolve as 
\begin{equation}
\frac{d Y_{\rm CO}}{dt} = n K_{\rm f}^{\rm CO} Y_{\rm C} Y_{\rm O} - K_{\rm d}^{\rm CO} Y_{\rm CO}\\
\end{equation}
\begin{equation}
\frac{d Y_{\rm SiO}}{dt} = n K_{\rm f}^{\rm SiO} Y_{\rm Si} Y_{\rm O} - K_{\rm d}^{\rm SiO} Y_{\rm SiO}\\
\end{equation}
where $Y$ is the number fraction of each element, $K_{\rm f}$ is the rate of formation due to radiative association, and $K_{\rm d} = K_{\rm th} + K_{\rm nth}$ is the rate of destruction with both thermal and non-thermal components.

Following \citet{Lazzati&Heger16}, CO forms by radiative association (${\rm C + O \rightarrow CO} + h \nu$) at a rate
\be
K_{\rm f}^{\rm CO} =  \frac{4.467 \times 10^{-17}}{\sqrt{  \left( \frac{T}{4467 \, {\rm K}} \right)^{-2.08} + \left( \frac{T}{4467 \, {\rm K}} \right)^{-0.22} }} \, \, \rm cm^3 s^{-1},
\ee
and is destroyed by thermal electrons at a rate 
\be
K_{\rm th}^{\rm CO} = K_{\rm ra} \left( \frac{h^2} {2 \pi \mu k_{\rm B} T} \right)^{-3/2} e^{\frac{-B_{\rm CO}}{k_{\rm B} T}}  \, \rm s^{-1},
\ee
where $h$ is Planck's constant, $\mu$ is the reduced mass of C and O, $k_{\rm B}$ is Boltzmann's constant, and $B_{\rm CO}$ is the binding energy of CO.

Similarly, SiO is formed by radiative association (${\rm Si + O \rightarrow SiO} + h \nu$)
at a rate (following \citealt{Cherchneff&Dwek2009})
\be
K_{\rm f}^{\rm SiO} = 5.52 \times 10^{-18} T^{0.31}  \, \, \rm cm^3 s^{-1}
\ee
 and destroyed by thermal electrons at a rate 
 \be
 K_{\rm th}^{\rm SiO} = 4.4 \times 10^{-10} e^{ \frac{-98600 \, {\rm K}}{T} } \, \, \rm s^{-1}.
\ee

The presence of shocks allows for another destruction pathway for molecules, particularly from the production of high energy particles (Appendix~\ref{sec:electrons}). Electrons accelerated at the shock, although subject to cooling via Coulomb scattering in the post-shock gas, can ionize molecules if they reach the central dust-forming shell with sufficient energy. Accelerated protons can collide with ambient protons in the post-shock gas, and a fraction of these events will produce pions ($\pi^{+/-}$) that decay to electron-positron pairs \citep{Kamae+06}.  Additionally, relativistic protons themselves may collide with and destroy molecules.  This total ionization rate is given as $K_{\rm nth} = {l_\gamma}/ {W_{\rm CO}} \, \rm s^{-1}$,
where $l_{\gamma}$ is the energy deposition rate per particle and $W_{\rm CO}$ is the mean energy per ionization.  The production rate of high-energy particles should coincide with the gamma ray emission and be proportional to the shock power (eq.~\ref{eq:Lsh2}).  We take $l_{\gamma}$ to be
\begin{equation}
\label{eq:Knth}
l_\gamma = \frac{\epsilon L_{\rm sh}}{N_{\rm tot}} f_{\rm CO} = \frac{\epsilon L_{\rm sh}  m_p}{M_{\rm ej} } f_{\rm CO}
\end{equation}
where $f_{\rm CO}$ incorporates the fraction of ejecta mass that is in CO molecules. We have introduced an efficiency factor $\epsilon$, corresponding to the fraction of the shock kinetic luminosity placed into the relativistic particles which are available to destroy molecules.  As summarized in Appendix~\ref{sec:electrons}, we estimate this efficiency to be $\epsilon = 10^{-4}$, a number we adopt in our fiducial models.  However, we also consider models with higher $\epsilon$ (Section \ref{sec:COdest}).

Our assumption of a carbon-neutral region to late times (Appendix \ref{sec:ionization}) allows us to neglect reactions involving ions. \citet{Pontefract&Rawlings04} found that neutral hydrogen reactions are important for forming CO at early times in nova outflows.  We neglect the impact of hydrogen because we find that radiative association is already sufficient to saturate CO (see Section~\ref{sec:results}). Certain H-reactions may also destroy CO at later times, but we estimate that their rates are negligible compared to the thermal and non-thermal destruction rates (e.g., \citealt{Cherchneff&Dwek2009} and references therein).

\subsection{Dust Nucleation and Growth}

Dust nucleation describes the process by which gas undergoes a phase transition from single atoms to more complex clusters. To capture this process precisely requires solving a complex reaction network that allows for the formation of small molecules, which eventually$-$potentially via many pathways$-$grow to macroscopic dust grains. It is generally intractable to follow such a complex process and well beyond the scope of this work.

Classical nucleation theory (CNT, \citealt{BeckerDoring35, Feder66}) simplifies this significantly by providing analytical expressions for the nucleation rate and the growth rate of grains. In CNT, when a gas becomes supersaturated, grains form as seed clusters that subsequently grow by accretion of other monomers. While CNT is considered an over-simplified approach for approximating grain formation (see \citealt{DonnNuth1985} and references therein), we utilize it here as a first step in exploring the possibility of grain formation within the post-shock gas. To gain insight into specific details of the grain formation process, one would need to consider more in depth approaches, such as chemical networks \citep{Cherchneff&Dwek2009} and/or atomistic nucleation models \citep{Mauney+2015}.

We track the formation of two different dust species --- carbon grains and forsterite (Mg$_2$SiO$_4$) to represent the silicate grain population. While observations often indicate the presence of both carbonaceous and silicate dust, the precise nature of the silicates is often unknown. At later times after the dust event when infrared spectra could shed light on the silicate species, the silicates often appear amorphous, lacking distinct spectral features. 

Following CNT, we implement dust nucleation as now described.  Carbon becomes supersaturated when its density exceeds the equilibrium density $n_{\rm eq}$, or the saturation ratio $S_{\rm C}$ exceeds unity,
\begin{equation}
S_{\rm C} = n_C^{\rm gas} / n_{\rm eq},
\end{equation}
where $n_C^{\rm gas}$ is the number density of free carbon in the gas state, and $n_{\rm eq} = (6.9 \times 10^{13} {\rm erg}) (k_B T)^{-1} \exp{(-84428.2 \, {\rm K} /T )} $ cm$^{-3}$ \citep{KeithLazzati}.
Silicate grains, rather than being built by single atoms, are composed of several species ($\rm 2Mg + SiO + 3O \rightarrow Mg_2SiO_4$). The saturation ratio depends on each of the abundances, and is given by
\begin{equation}
\ln{S_{\rm Si}} = - \frac{\Delta G}{k_B T_{\rm cs}} + \sum_i \nu_i \ln{p_i},
\end{equation}
where $T_{\rm cs}$ is the temperature of the cold shell of gas $\Delta G$ is the Gibbs free energy of the reaction, $\nu_i$ are the stoichiometric coefficients, and $p_i$ are the partial pressures of each species \citep{KozasaHasegawa87, Nozawa2003}.

If the saturation is greater than unity, nucleation occurs exponentially at a rate (\citealt{Feder66})
\begin{equation}
J_i = \sqrt{ \frac{ c_{\rm sh}^3 \sigma_i v_i^2} { 18 \pi^2 m_i}} n_i^2 
\exp {\left( \frac{-4 \, c_{\rm sh}^3 \, v_i^2 \, \sigma_i^3}{27 (k_B T)^3 (\log{S_i})^2 } \right) } \, \, \rm cm^{-3} s^{-1}.
\end{equation}
For carbon grains, $\sigma_i$, $v_i$, $m_i$, $n_i$, and $S_i$ correspond to the specific surface energy, volume, mass, number density, and saturation ratio of carbon, respectively. For forsterite grains, the monomers are not present in the gas phase, so we adopt the \emph{key species} approach, as introduced in \citealt{KozasaHasegawa87}, in which the nucleation rate is governed by the single species that sets the slowest nucleation rate (i.e. the one with the lowest collisional frequency). Hence the rate is dependent on the abundance of Mg or O, depending on which is less abundant, and $\sigma_i$, $v_i$, $m_i$ correspond to the specific surface energy, volume, and mass of the monomer. In this work $c_{\rm sh}$ refers to the grain shape factor\footnote{In other works the shape factor is often denoted as $c_{\rm s}$, a term we avoid so as to prevent confusion with the gas sound speed.}, which we set to $c_{\rm sh} = (36 \pi)^{1/3}$ assuming spherical grains.

For each species $i$, grains form at a critical size given by 
\begin{equation}
N_{\rm crit} = \frac{8 \, c_{\rm sh}^3 \, v_i^2 \, \sigma_i^3}{27 (k_B T \log S_i)^3},
\end{equation}
and we impose a minimum size of 2 monomers.  Any previously formed grains will continue to grow by accreting at a rate
\begin{equation}
\frac{dN}{dt} = n_i^{\rm gas} c_{\rm sh} (N_0 v_i)^{2/3} \sqrt{\left( \frac{k_B T}{2 \pi m_i} \right)} \, \, \rm s^{-1},
\end{equation}
where $N_0$ is the current size, $v_i$ is the volume of either solid C or  Mg$_2$SiO$_4$, and $m_i$ is the mass of the molecule. 

In cases where the radiation field is strong, grains will easily become hotter than the surrounding gas, potentially leading to evaporation.  Here we assume that the dust temperature is the same as that of the gas, and we neglect evaporation since we are considering a region that is optically thick to the UV radiation field.

\subsection{Elemental Abundances}
\label{sec:abundances}

\citet{Evans&Gehrz08} overview the abundances in nova ejecta.  The TNR results in the ejecta being enriched in CNO elements (e.g.~\citealt{Starrfield+98}), and heavier elements like magnesium and silicon can also have their abundances significantly altered by proton capture processes (e.g.~\citealt{Nofar+91}).  Spectroscopic observations suggest that a significant fraction of the ejecta is likely to be dredged up material from the white dwarf interior (e.g.~\citealt{Ferland&Shields78}), also enhancing the C/O/Ne abundances.  There is considerable uncertainty in observationally-determined abundance measurements, which depend on modeling the nebular phase and accounting for difficulties such as extinction (both local and interstellar), clumpiness, the incident source of ionizing radiation, and geometry (\citealt{Schwarz+97}, \citealt{Schwarz+02}, \citealt{Schwarz+07}).  In our case, the latter uncertainty relates to whether the portion of the ejecta which is measured spectroscopically the same as that being shocked.  The line shapes of emission lines indicate that in some cases, oxygen and Ne-rich ejecta may reside in different parts of the ejecta (\citealt{Helton+12}).  

We focus on CO novae, which are more frequently dust producers than ONe novae.  For our fiducial model, we adopt ejecta abundances matching those of the CO nova with the most comprehensive photo-ionization modeling to date, namely GQ Mus.  \citet{Morisset&Pequignot96} find abundances of $X_{\rm H} = 0.372$, $X_{\rm He} = 0.394$, $X_{\rm C} = 8 \times 10^{-3}$, $X_{\rm N} = 0.125$, $X_{\rm O} = 9.5 \times 10^{-2}$, $X_{\rm Mg} = 6.7 \times 10^{-4}$, and $X_{\rm Si} = 7.7 \times 10^{-4}$.  Note that the C:O ratio is very low for these abundances, the consequences of which are discussed in the following section.

\begin{table*}
\begin{center}
{Parameters/Results}
\vspace{2mm}
\begin{tabular}{ccccccccccc}
\hline\midrule
\multicolumn{1}{c}{Trajectory} &
\multicolumn{1}{c}{C:O}&
\multicolumn{1}{c}{$\epsilon$} &
\multicolumn{1}{c}{$t_{\rm cond}^{\rm solid C}$} &
\multicolumn{1}{c}{$t_{\rm cond}^{\rm Mg_2SiO_4}$} &
\multicolumn{1}{c}{$n_{\rm cond}^{\rm solid C}$} &
\multicolumn{1}{c}{$n_{\rm cond}^{\rm Mg_2SiO_4}$} &
\multicolumn{1}{c}{$\bar{a}_{\rm solid C}$} &
\multicolumn{1}{c}{$\bar{a}_{\rm Mg_2SiO_4}$} &
\multicolumn{1}{c}{$X_{\rm solid C}$} &
\multicolumn{1}{c}{$X_{\rm Mg_2SiO_4}$} \\
\cmidrule{1-11}
\vspace{-0.2cm}
\\
  & - & - & days & days & $\rm cm^{-3}$ & $\rm cm^{-3}$ & $\rm \mu m$ & $\rm \mu m$ & - & -\\
\hline
\\
\multirow{1}{*}{Uniform sphere } 
& $0.11$ & $10^{-4}$ & $5.1$ & $11.5$ & $5 \times 10^9$ & $5 \times 10^8$  
& $6 \times 10^{-4}$ & $3 \times 10^{-4}$ & $0.008$ & $6 \times 10^{-4}$ \\  \midrule
\multirow{9}{*}{Shock compression} 
&  \multirow{3}{*}{$0.11$} 
& $10^{-4}$  & $55.7$ & $2.6$ & $2 \times 10^{13}$  & $5 \times 10^{14}$ 
& $2 \times 10^{-4}$ & $0.4$ & $2 \times 10^{-12}$  & $0.002$ \\
&  & $10^{-3}$ & $41.2$ & $2.6$ & $6 \times 10^{13} $ & $5 \times 10^{14}$ 
& $2 \times 10^{-4}$ & $0.4$ &  $4 \times 10^{-10}$ & $0.002$   \\ 
& & $10^{-2}$ & $32.9$ &  $2.6$ & $1 \times 10^{14}$ & $5 \times 10^{14}$ 
& $5 \times 10^{-4}$ & $0.4$  &  $9 \times 10^{-8}$ & $0.002$ \\   \cmidrule{2-11} 
& \multirow{3}{*}{$0.56$} 
& $10^{-4}$ & $34.9$ & $4.4$ & $8 \times 10^{13}$ & $5 \times 10^{14}$ 
& $3 \times 10^{-4}$ & $0.3$ & $2 \times 10^{-8}$  & $0.002$  \\
& & $10^{-3}$ & $28.3$ & $4.4$ & $1 \times 10^{14}$ & $5 \times 10^{14}$ 
& $0.001$ & $0.3$ &  $7 \times 10^{-7}$ & $0.002$ \\ 
& & $10^{-2}$ & $23.0$ & $4.4$ & $2 \times 10^{14}$ & $5 \times 10^{14}$ 
& $0.008$ & $0.3$ & $1 \times 10^{-5}$ & $0.002$ \\  \cmidrule{2-11} 
& \multirow{3}{*}{$1.12$} 
& $10^{-4}$ & $7.6$ & $19.0$ & $4 \times 10^{14}$ & $2 \times 10^{14}$  
& $8.3$ & $0.07$ & $0.009$ & $0.0005$ \\
& & $10^{-3}$ & $7.6$ & $18.7$ & $4 \times 10^{14}$ & $2 \times 10^{14}$ 
& $8.5$ & $0.07$ &  $0.009$ & $0.0006$ \\  
& & $10^{-2}$ & $7.6$ & $17.5$ & $4 \times 10^{14}$ & $2 \times 10^{14}$ 
& $8.6$ & $0.1$ &  $0.009$ & $0.002$ \\  
\hline
\hline
\end{tabular}
\end{center} 
\caption{Summary of results for different trajectories. C:O is the ratio of initial carbon to oxygen by number, $\epsilon$ is the efficiency of CO destruction by non-thermal particles, $t_{\rm cond} = t - t_{\rm shock}$ is the time grains begin to form after the post-shock fluid has entered the cold shell, and $n_{\rm cond}$ is the density at this time. $\bar{a}$ is the mean size of the grains weighted by mass after the growth has saturated (almost immediately). $X$ is the final mass fraction of each grain.\label{table:results}
}
\end{table*}

\section{Results}
\label{sec:results}
Here we describe in detail the dust formation in our fiducial model before discussing the results from variations about this model. The results are summarized in Table~\ref{table:results}, where for each run we include the dust condensation time, the density at this time, the average grain sizes, and the total dust mass fraction.  

We quote times as measured with respect to when a typical fluid element enters the shock, which, based on the observed gamma-ray emission, is typically weeks after the initial outburst itself.  Even though the post-shock gas is shielded from the ionizing radiation of the WD, it is still heated by the optical radiation field reprocessed through the gas.  As a result of this temperature floor (eq.~\ref{eq:Trad}), dust cannot condense until the fluid element reaches sufficiently large radii, at a time which depends on the outflow velocity. In this study we consider a single value of the ejecta velocity for simplicity.  However, given the observed diversity of velocities, both between different novae and within individual events, dust formation can commence across a range of timescales relative to the start of the outburst (e.g.~\citealt{Evans&Gehrz12}).  

Our fiducial model makes the conservative assumptions of a temperature floor set by the central WD (eq.~\ref{eq:Trad}) and a low efficiency of CO destruction, $\epsilon = 10^{-4}$.  However, even given these assumptions, we find that molecule and dust formation are efficient.  Reasonable deviations from these adopted parameters generally serve only to increase the total dust yield.  

\begin{figure}
\begin{center}
\large{}\par\medskip
\subfloat{
\includegraphics[width=.5\textwidth]{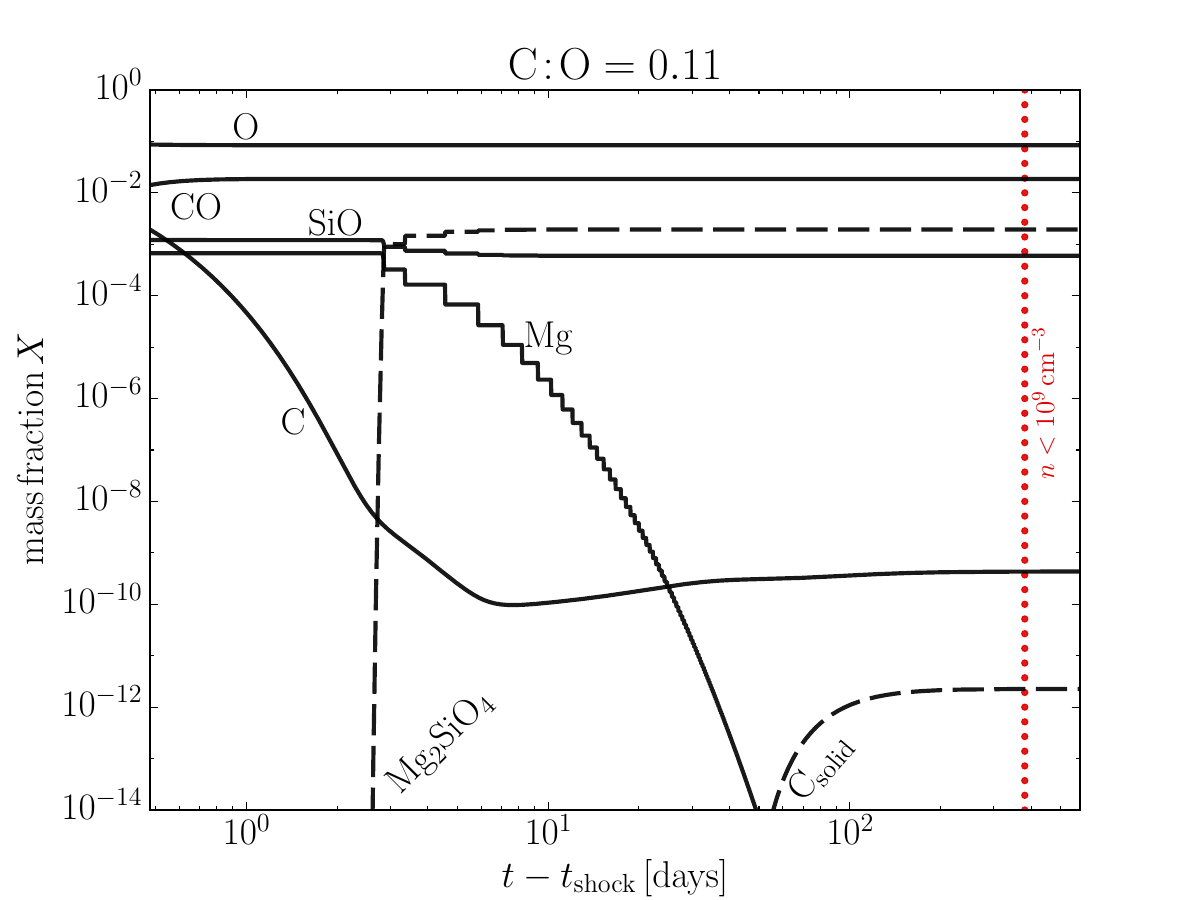}}

\vspace{-4mm}

\subfloat{
\includegraphics[width=.5\textwidth]{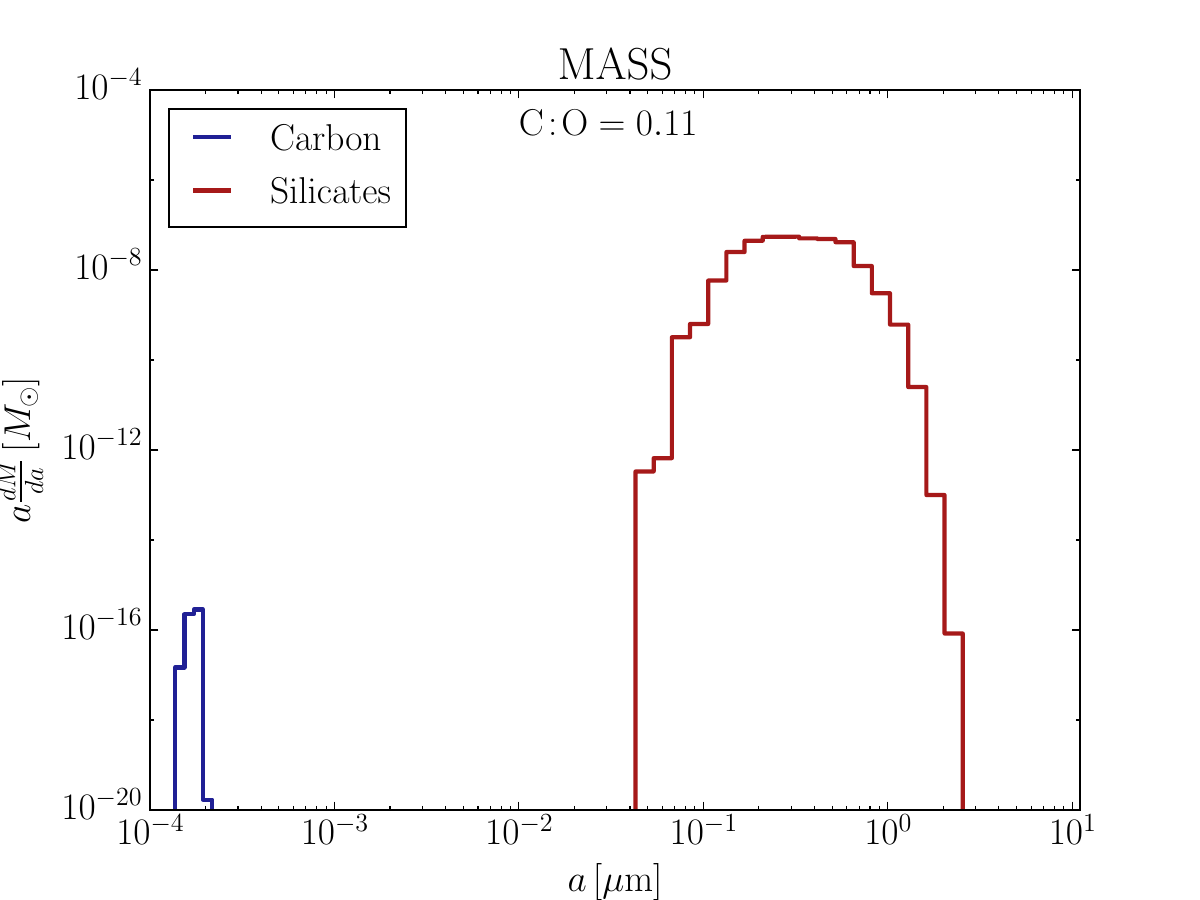}}

\vspace{-4mm}

\subfloat{
\includegraphics[width=.5\textwidth]{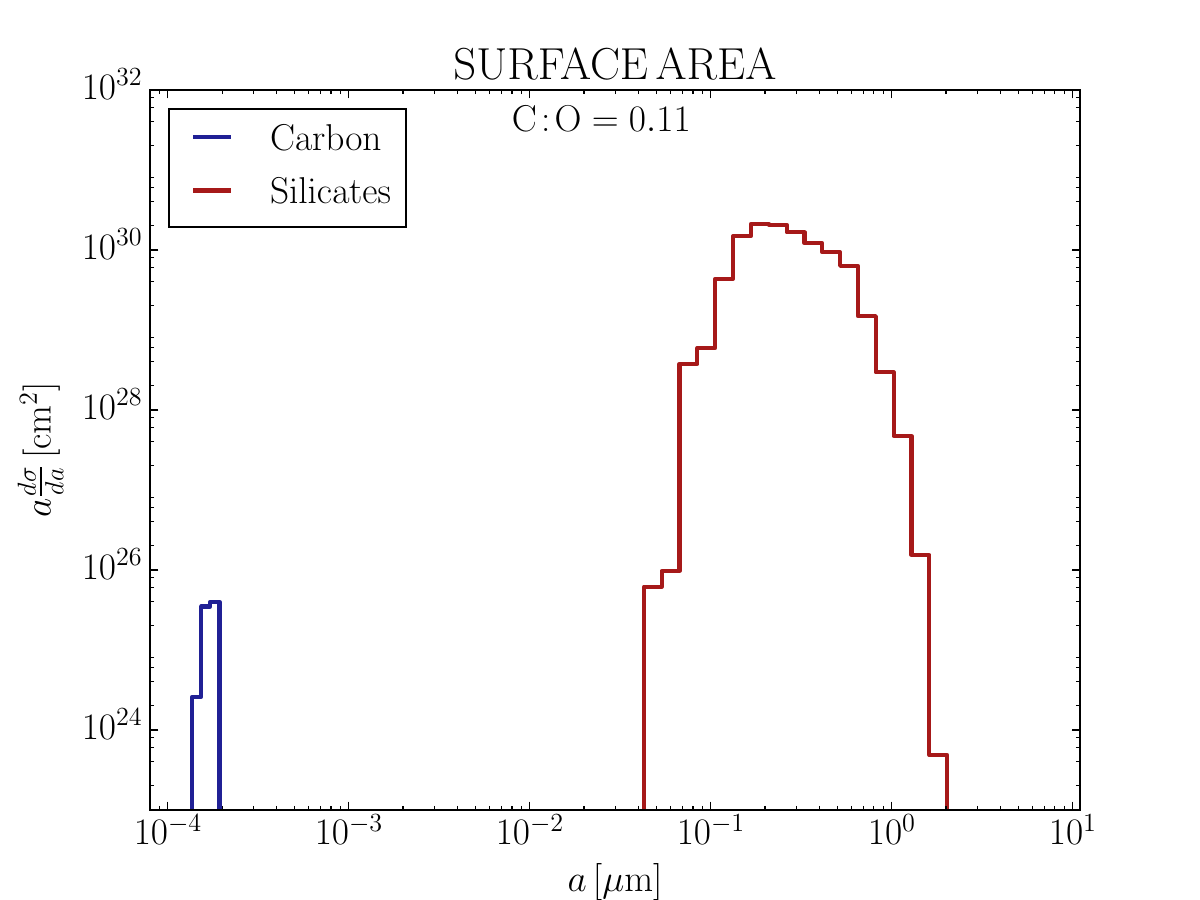}}

\caption{Top: Chemical evolution of gaseous elements and dust. CO saturates at early times, reducing the availability of carbon for forming grains. Silicates ($\rm Mg_2SiO_4$) form rapidly, and a trace amount of carbon ``grains'' ($\rm C\, solid$) condense at $t \sim 2 \, \rm mo$. The red dotted line indicates at what time the density decreases below $n \le 10^9 \, \rm cm^{-3}$, at which point we expect harsh radiation to penetrate the fluid. Middle: The resulting log-normal mass distribution of carbon (blue line) and silicates (red line), normalized for a total shocked ejecta mass $M_{\rm ej}=10^{-4}$. Silicate grains dominate the mass with a characteristic size of $a \approx 0.3 \, \mu \rm m$. Bottom: Surface area distribution. The contribution to the surface area by silicates alone is sufficient to cause visual extinction with $\tau \sim 1$.  }
\label{fig:fid_X_mass}
\end{center}
\end{figure}

The top panel of Figure~\ref{fig:fid_X_mass} shows the chemical evolution by mass fraction as the fluid parcel is compressed, cools, and then expands.  The high densities achieved due to shock-induced compression allow CO to quickly saturate and deplete all free carbon, preventing the condensation of carbon grains. Oxygen and magnesium are abundant, however, and within days of reaching peak density there is a burst of silicate grain formation. 

Rather than being a gradual process, the first burst of nucleation occurs rapidly. It is efficient enough to use all available monomers, leaving none for subsequent grain growth. Grain growth after this initial burst occurs more gradually as CO is destroyed (freeing carbon and oxygen) at later times; however, the change in grain size is negligible in our fiducial model due to the weak level of CO destruction (low $\epsilon$).  This burst-like nature of condensation results in a log-normal size distribution that is dominated by a representative size, as shown by the mass distribution in the middle panel of Figure~\ref{fig:fid_X_mass}, which we have normalized to an assumed total mass of shocked gas of $10^{-4} M_{\odot}$, comparable to the total ejecta mass (eq.~\ref{eq:Lsh} and surrounding discussion).  The majority of the dust mass is contained in silicate grains ($X_{\rm Mg_2SiO_4}\sim10^{-3}$) with radii $a \sim 0.4 \, \mu m$.  

Our fiducial abundances are very carbon-poor: the ratio of $Y_{\rm C}/Y_{\rm O}$, denoted as C:O, is less than 1.  The formation of carbon grains is therefore dependent on the destruction of CO by non-thermal processes. This occurs at later times as the density decreases, resulting in the delayed condensation of solid carbon ($t \sim 60$ days) with the mass distribution peaking at very small sizes ($a \sim 10^{-4} \, \mu  m$).  Note that by our definition, a grain must consist of at least 2 monomers (corresponding to minimum radii $a_{\rm min} \approx \! 5 \times10^{-4} \mu$m for carbon grains, and $a_{\rm min}\approx 9 \times 10^{-4} \mu$m for silicates). Hence grains with sizes less than $a\lesssim 10^{-4}$ are not necessarily physically meaningful-- rather they are an artifact of low-efficiency grain formation from CNT.
Nevertheless, we note that in the presence of a UV radiation field, these smaller grains would heat up due to interactions with photons and likely not survive (e.g.~\citealt{Kochanek14}).  However, given the high densities of the post-shock gas, small grains are protected in a region that is optically thick to the harsh radiation, allowing them to survive and continue to grow if the necessary elements are available. The time at which the density decreases to the value below which the carbon would be ionized ($n_{\rm ion}$; eq.~\ref{eq:nion}) is denoted as a vertical dotted line in Figure~\ref{fig:fid_X_mass}.  Ultimately, in our fiducial model, the CO destruction is not efficient enough to allow for significant carbon grain formation. 

The bottom panel of Figure~\ref{fig:fid_X_mass} shows the surface area distribution of the grains, a quantity important for determining the magnitude of optical extinction for viewers along the line of sight of the shocks.  As a point of reference, for large grains with sizes in the geometric optics limit ($a \gtrsim \lambda$, where $\lambda$ is the observing wavelength), to reach an optical depth $\tau_{\lambda} \gtrsim 1$ requires the surface area of the grains at $t \sim$ 1 month to be $\sigma \gtrsim 4 \pi f_{\Omega} R_{\rm sh}^2 \approx 4 \pi f_{\Omega} (v t)^2\sim10^{29} \rm \, cm^2$ for a shell velocity of $v \approx 1000 \rm \, km \, s^{-1}$ and order-unity shock covering fraction $f_{\Omega} \sim 1/3$.  In our fiducial run, the surface area from large silicate grains alone is therefore enough to produce several magnitudes of complete optical extinction (see also Section \ref{sec:discussion}). 

A unique feature of dust formation in post-shock gas of radiative shocks in novae (e.g., as compared to dust formation in lower density astrophysical environments, or even in supernovae) is that the condensation temperature is reached at extremely high densities. Not only does this justify our assumption of a carbon-neutral region, but$-$given the proper abundances$-$grain formation is remarkably efficient.  On one hand, such efficient condensation can explain the observed rapid growth of grains in some novae (e.g. V705 Cas, \citealt{Shore+94}).  On the other hand, such high densities make CO formation too efficient, which in an oxygen-rich environment (C:O$\ll$1) makes it challenging to simultaneously produce carbon grains.  As we now discuss, increasing the efficiency of carbon formation from that of our fiducial model requires either considering regions of the ejecta which are more carbon-rich (Section~\ref{sec:COratio}), or higher efficiencies of molecule destruction by non-thermal particles (Section~\ref{sec:COdest}).

\subsection{Effect of a higher C:O ratio}
\label{sec:COratio}

To consider dust formation in carbon-rich parts of the ejecta, we run otherwise identical models with enhanced carbon abundance (leaving the oxygen abundance the same as the fiducial model).  The results of these models are summarized in Table~\ref{table:results}.

First, by increasing the value of $X_{\rm C}$ by a factor of 5 (renormalizing the other abundances accordingly), we find the same qualitative behavior as in the fiducial case, albeit with a slightly higher carbon dust yield.  This is not surprising because the C:O ratio is still less than unity for this model (C:O = $0.56$). 

\begin{figure}
\begin{center}
\large{}\par\medskip
\subfloat{
\includegraphics[width=.5\textwidth]{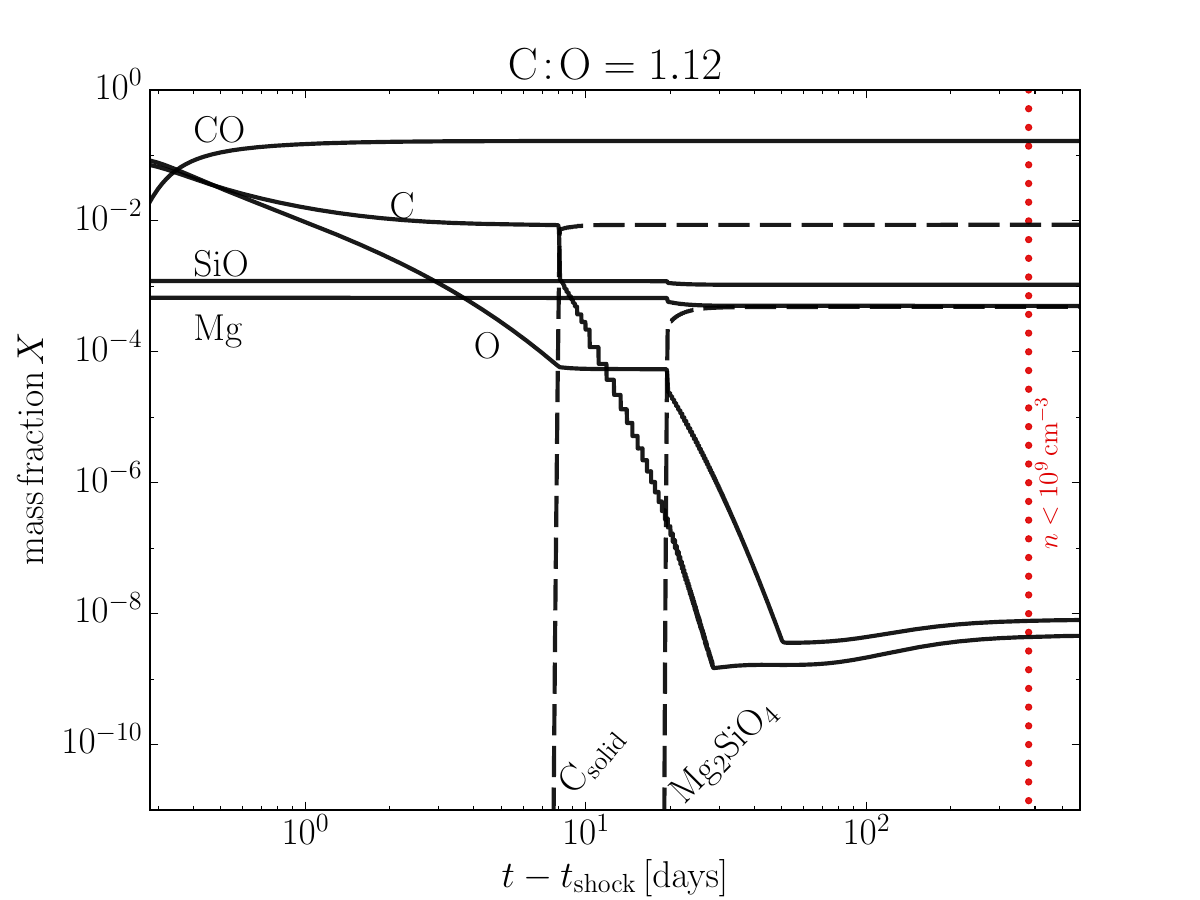}}

\vspace{-4mm}

\subfloat{
\includegraphics[width=.5\textwidth]{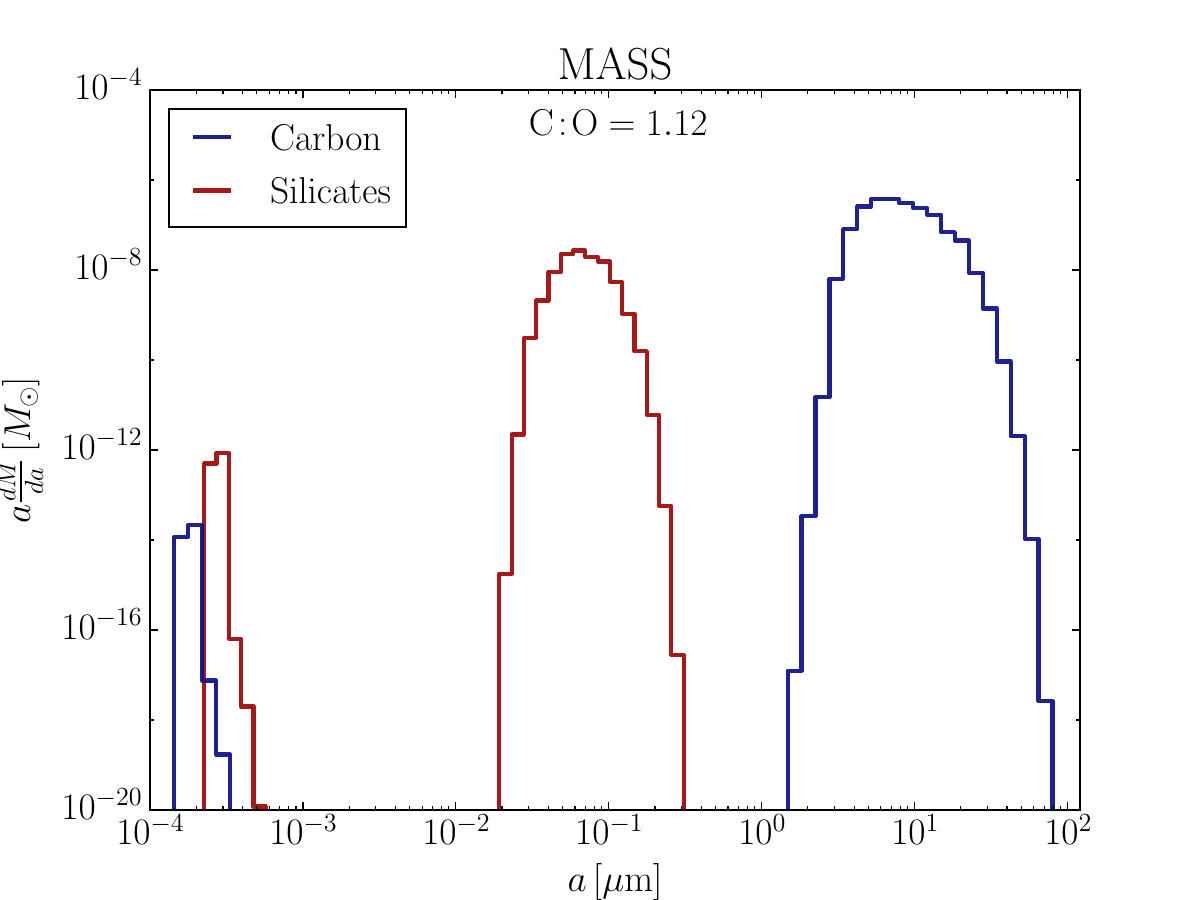}}

\vspace{-4mm}

\subfloat{
\includegraphics[width=.5\textwidth]{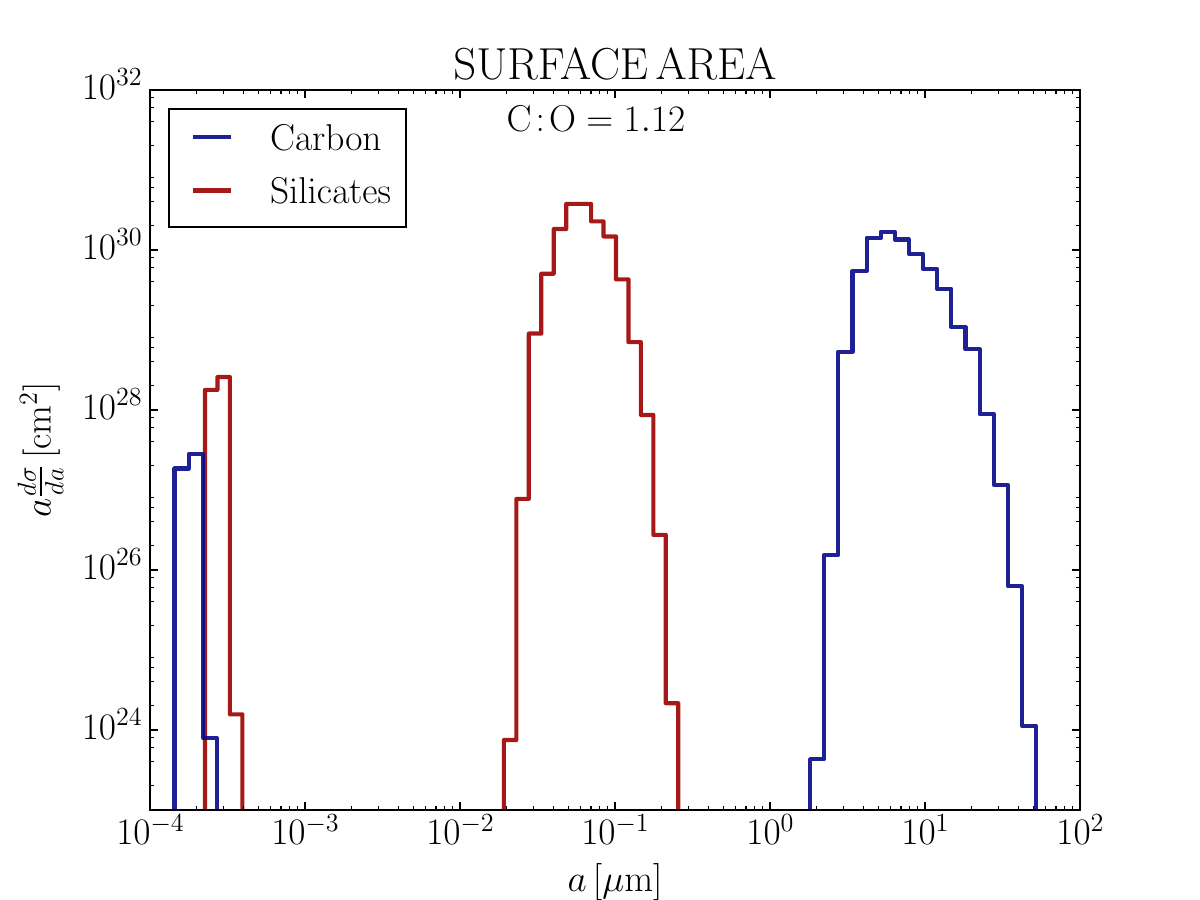}}

\caption{Top: Chemical evolution of gaseous elements and dust for C:O $>$ 1. Both carbon and silicate grains condense efficiently.
Middle: Mass distribution of carbon (blue line) and silicate (red line) grains (normalized to $M_{\rm ej}=10^{-4}$). Carbon grains dominate in mass with an average mass-weighted size of $\bar{a}\sim 8 \mu$m, and silicates peak at $\bar{a}\sim0.1 \mu$m. The second peak at smaller sizes manifests from a second, weaker burst of nucleation at later times, as CO destruction releases carbon and oxygen.
Bottom panel: Surface area distribution.  Both carbon and silicate grains contribute amply to optical extinction. }
\label{fig:CO_ratio}
\end{center}
\end{figure}

More interestingly, by increasing the value of $X_{\rm C}$ by a factor of 10 to bring the C:O ratio above unity (C:O = $1.12$), we find the efficient formation of both carbon and silicate grains. Figure~\ref{fig:CO_ratio} shows the time evolution of dust mass fraction for this case, as well as the mass and surface area distributions for comparison to the fiducial model (Fig.~\ref{fig:fid_X_mass}). The excess carbon which is available following CO formation condenses into grains prior to silicate formation, as we expect given the higher carbon dust condensation temperature.  The total mass of carbon grains peaks at the characteristic size of $a \sim 8 \mu$m.  We do not take such large dust grains literally, as CNT tends to overestimate the formation rate of grains \citep{Mauney+2015,Lazzati&Heger16}.  However, our results make certain that nucleation is rapid and growth is efficient, resulting in large grains.  

Silicates form soon after carbon dust, reaching sizes around $a \sim 0.1 \mu m$.  While silicates contribute less to the total dust mass ($X_{\rm Mg_2SiO_4}\sim10^{-4}$), their formation is not significantly affected by the C:O ratio despite the fact that oxygen monomers can be locked up in CO.  Rather, the formation of silicates is more strongly dependent on the abundance of Mg. The contribution of silicates to the total mass and surface area is still significant, in the particular the potential for significant visual light dust extinction along the line of sight of the shocks.

\subsection{Efficiency of CO destruction}
\label{sec:COdest}

In our models explored thus far, the efficiency of CO destruction was fixed to $\epsilon = 10^{-4}$, for which the late-time formation and growth of dust enabled by CO destruction is relatively modest. However, the value of $\epsilon$ is uncertain, as it depends on the efficiency of leptonic versus hadronic acceleration at the shock, the shock velocity, and on uncertainties in the cross-section for destroying CO by relativistic electrons (see Appendix~\ref{sec:electrons}).  In this section we consider otherwise identical models to the fiducial carbon-poor case (C:O = 0.11), but with higher destruction efficiencies of $\epsilon = 10^{-3}, 10^{-2}$ (Table~\ref{table:results}). 

\begin{figure}
\subfloat{
\includegraphics[width=0.5\textwidth]{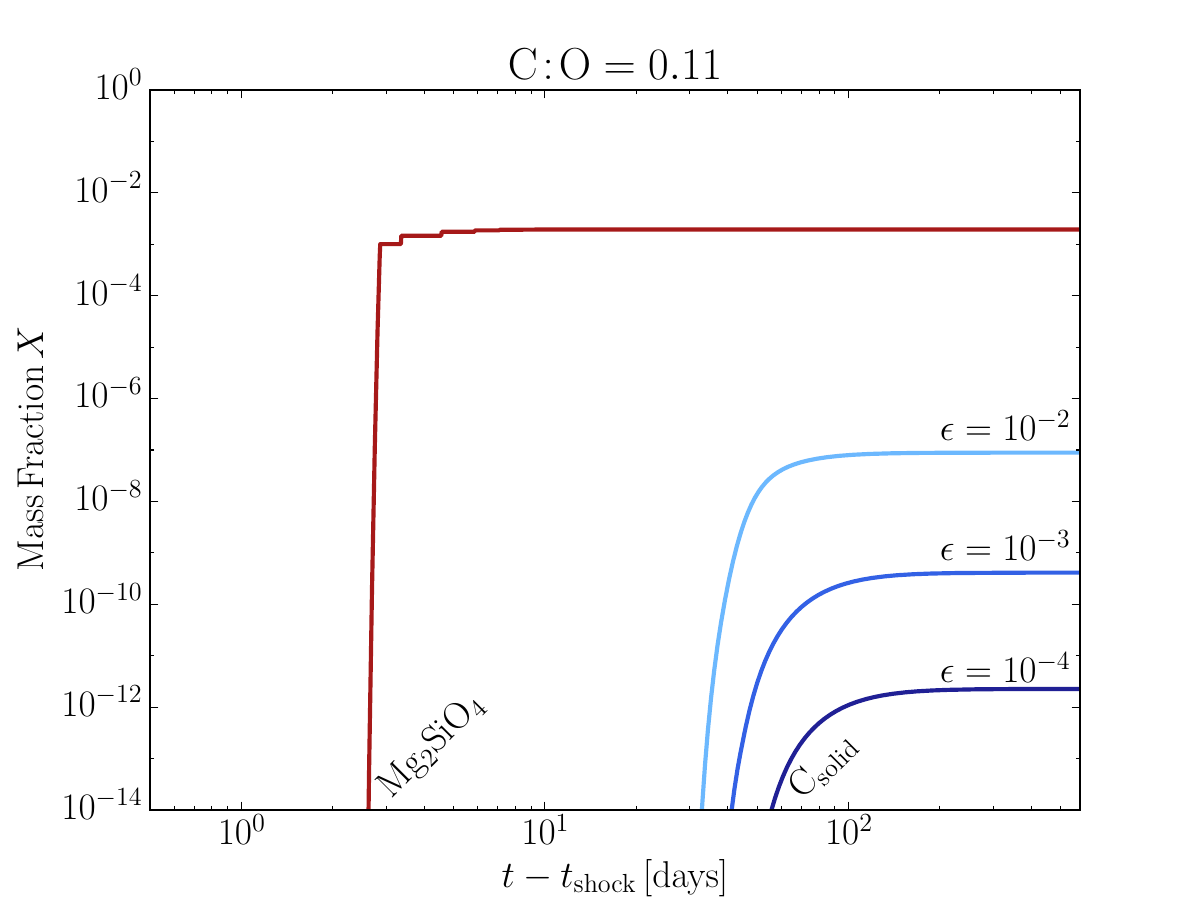}}

\vspace{-4mm}

\subfloat{
\includegraphics[width=0.5\textwidth]{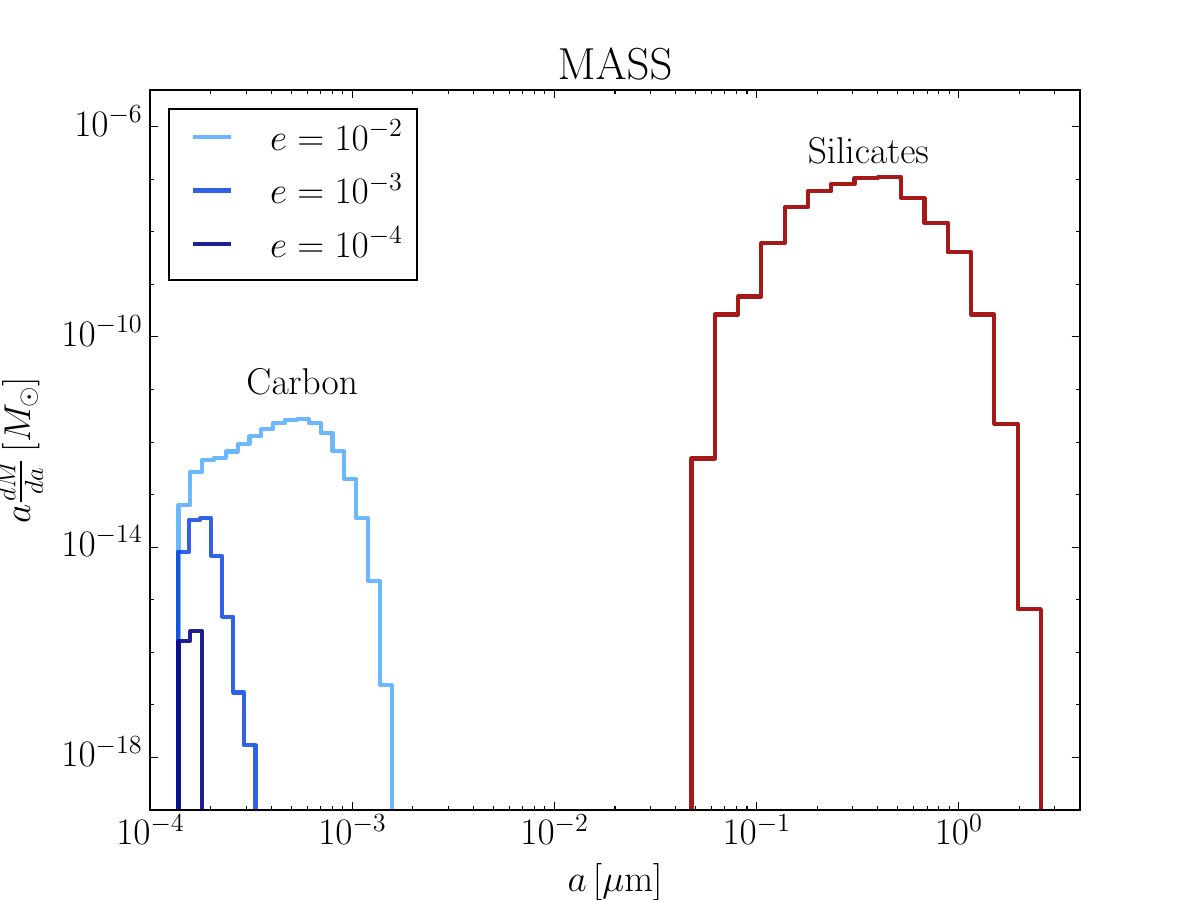}}

\vspace{-4mm}

\subfloat{
\includegraphics[width=0.5\textwidth]{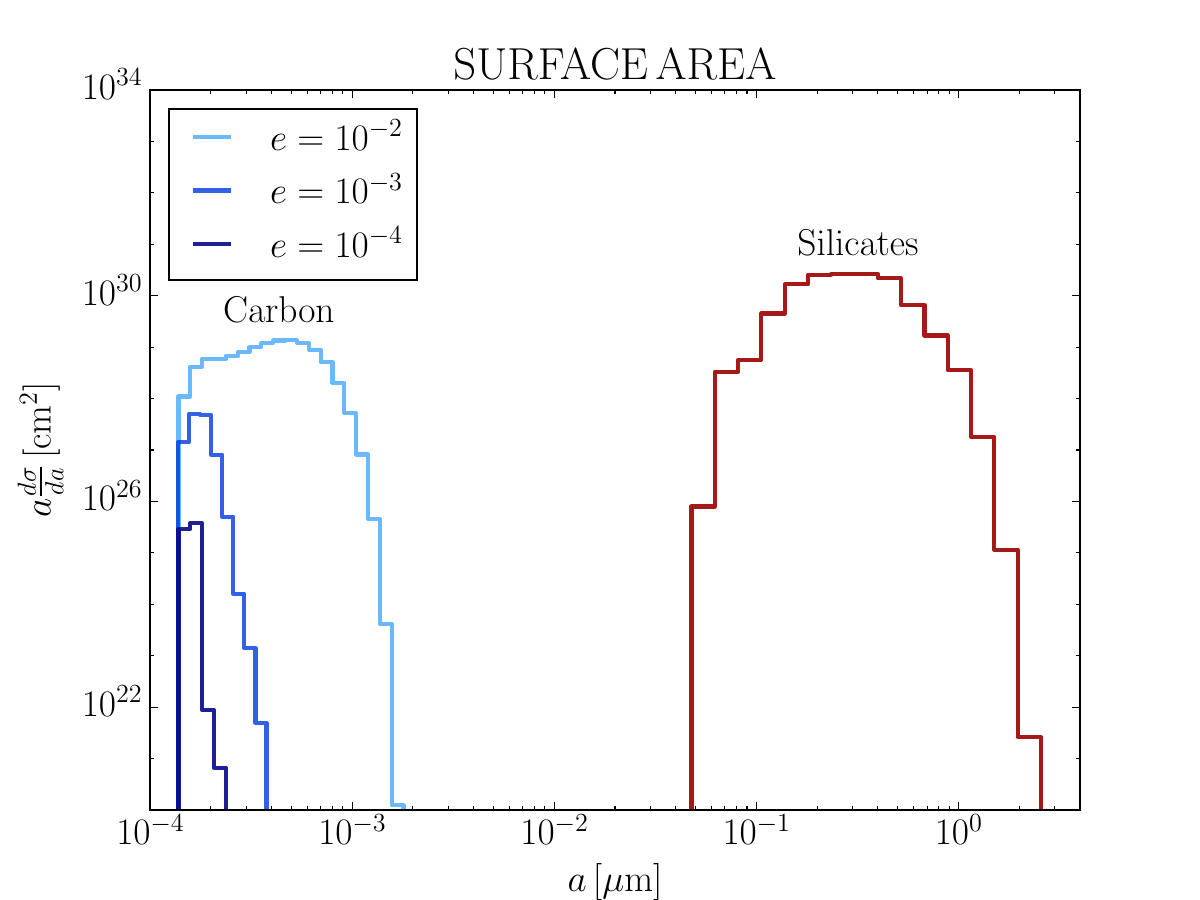}}
\caption{ Top: Chemical evolution of carbon and silicate grains, where we have varied the efficiency of molecular destruction $\epsilon$. For higher values of $\epsilon$ (lighter blue lines) carbon grains condense earlier and the total mass of solid carbon increases considerably. 
Middle: Final mass distribution of both grain types as we increase $\epsilon$. For higher $\epsilon$ (lighter blue lines), carbon grains are able to reach larger sizes. Silicate formation is relatively unaffected.
Bottom: Surface area distribution.  At the highest estimate of $\epsilon = 10^{-2}$, the smaller carbon grains contribute considerably to the total surface area. }
\label{fig:change_eps}
\end{figure}

Figure~\ref{fig:change_eps} compares the dust formation model as we increase $\epsilon$.   Unsurprisingly, more efficient CO destruction leads to increasingly larger grains.  For $\epsilon = 10^{-2}$, the total mass fraction of carbon dust increases by several orders of magnitude ($X_{\rm solidC} \sim 10^{-7}$), with grain sizes reaching up to $a \sim 10^{-3} \mu \rm m$. This difference is more pronounced in the surface area distribution, where the carbon grains, although still much smaller than the silicates in terms of size and total mass, contribute comparably to the total surface area.  However, given their smaller sizes $a \ll \lambda$, they will still contribute significantly less visual extinction than the silicate grains.  We conclude that, although CO destruction by non-thermal particles enhances the abundance of carbon grains in oxygen-rich ejecta, it is unclear whether this alone is sufficient to explain the simultaneous detection of carbonaceous and silicate grains in individual novae.

\subsection{The need for ejecta over-densities}
\label{sec:overdensities}
 Even if shocks or other inhomogeneities were not present in nova ejecta, the densities might at first seem  high enough for dust formation to occur.  However, here we demonstrate the necessity of significant density enhancements (e.g. by radiative shock compression).  We do this by showing for comparison dust formation models calculated for a parcel of gas expanding as a part of a spherical homogeneous outflow of total mass $10^{-4}M_{\odot}$ (the solid grey line trajectory in Figure~\ref{fig:thermo}), i.e. an approximation to the expected homologous state of the ejecta at the dust formation epoch absent internal shocks. 

\begin{figure}
\begin{center}
\large{}\par\medskip
\subfloat{
\includegraphics[width=.5\textwidth]{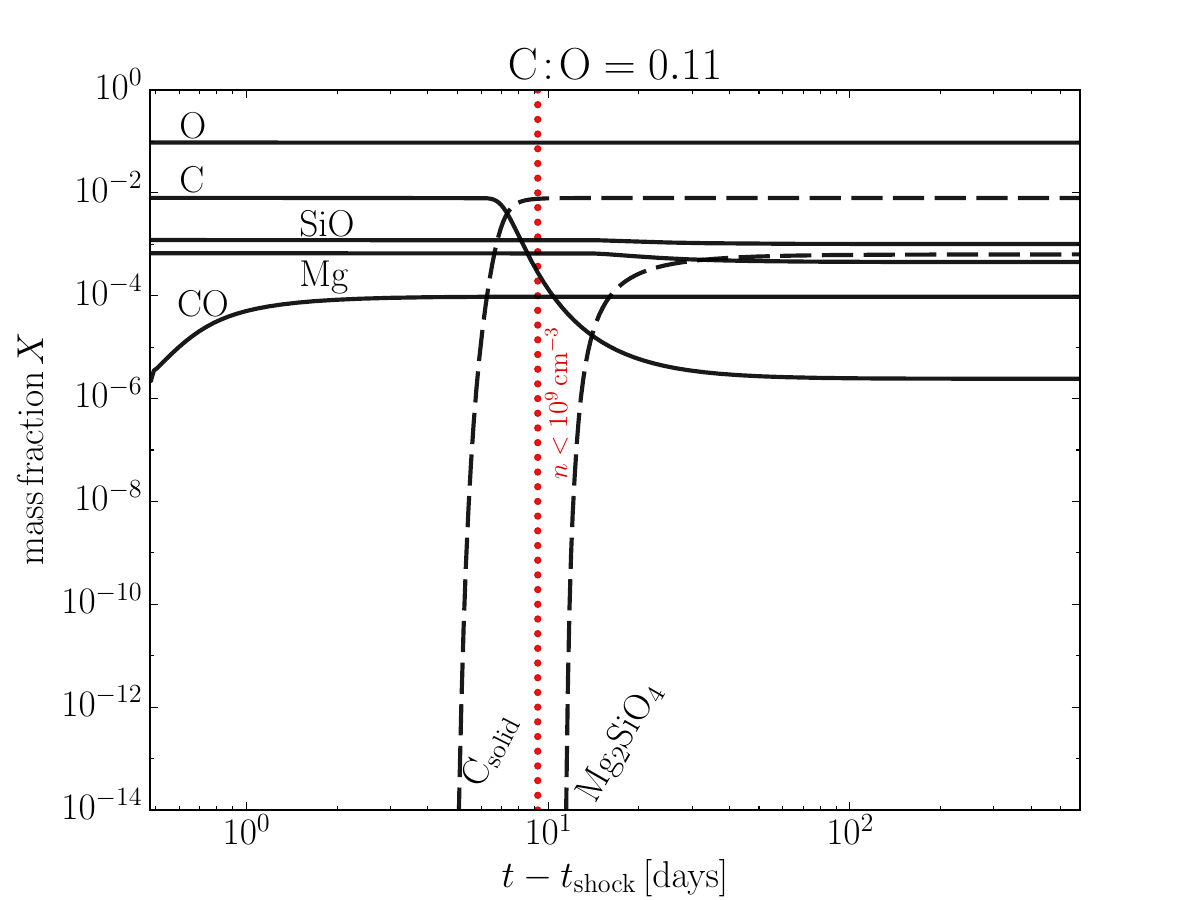}}

\vspace{-4mm}

\subfloat{
\includegraphics[width=.5\textwidth]{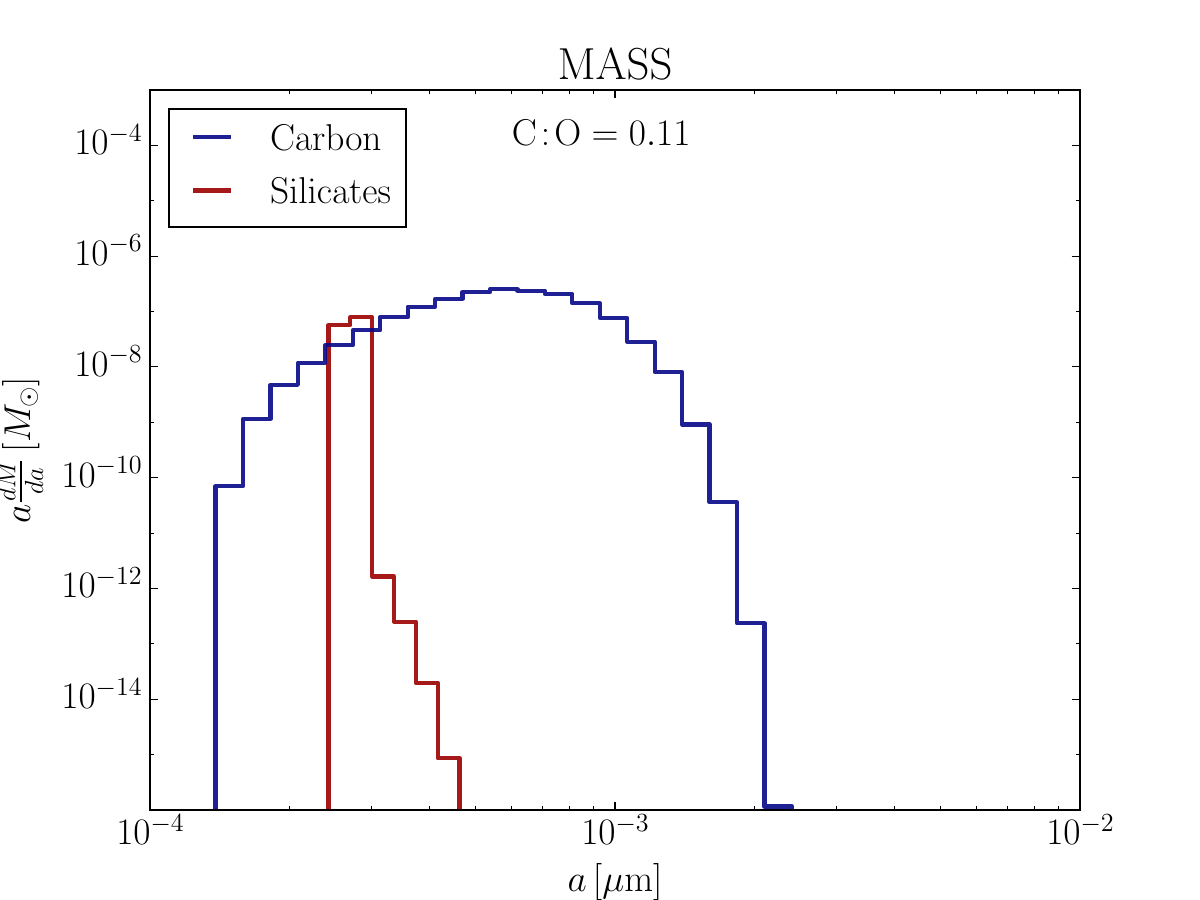}}

\caption{Top: Chemical evolution of gaseous elements and dust during the expansion of a uniform sphere. Left of the dotted vertical line, the density is above values at which all elements remain neutral ($n > 10^9 \, \rm cm^{-3}$). In this trajectory, CO does not saturate which allows for the formation of both carbon and silicate grains. However, grains form at lower densities, at which we need to worry about certain elements (particularly carbon) being photo-ionized. 
Bottom: Mass distribution of carbon (blue line) and silicates (red line) (normalized to $M_{\rm ej}=10^{-4}$). The mass is dominated by grains that are smaller than $a < 10^-3 \mu \rm m$. At such small sizes these grains are subject to evaporation by an external radiation field, particularly because they form in a lower density gas. These grains would likely not survive to contribute to any visual extinction.  }
\label{fig:sphere_X}
\end{center}
\end{figure}

Figure~\ref{fig:sphere_X} shows the chemical evolution of the spherical homogeneous trajectory.  Although grains condense when the temperature reaches $T \lesssim 1700$ K, their mean sizes of $a \sim 10^{-4}-10^{-3} \mu \rm m$ are significantly smaller than in the shock-compressed models and as inferred by observations of some novae.  Two other stark differences characterize the formation of dust with this trajectory.  First, because of the lower densities, CO does not approach saturation. Hence the non-thermal destruction of CO (even were shocks present) plays no critical role here because free carbon and oxygen remain abundant. 

The second difference, which is problematic for whether grains can form at all, is that the ejecta density at which grains condense is lower than in shock-compressed cases by at least 4 orders of magnitude (a pink dotted line in Fig.~\ref{fig:sphere_X} indicates the time when the density reaches $n = 10^{9} \, \rm cm^{-3}$).  This is also illustrated in the bottom panel of Figure~\ref{fig:thermo}, where symbols on the thermodynamic trajectories indicate the time of carbonaceous and silicate grain formation in different models.  In the spherically homogeneous case, solid carbon grains condense at $n_{\rm cond} \approx 5 \times 10^{9} \rm \, cm^{-3}$, followed by silicates at $n_{\rm cond} \approx 5 \times 10^8 \, \rm cm^{-3}.$  These values are close to, or less than, the critical density $n_{\rm ion} \approx 10^{9}$ cm$^{-3}$ (eq.~\ref{eq:nion}) below which critical elements (e.g. carbon or oxygen) will become photo-ionized by the central WD, hampering dust nucleation.  Even if dust formation could occur in such an environment, UV radiation from the WD would render such small grains vulnerable to evaporation.  We again conclude that extreme ejecta over-densities, such as those which arise naturally behind radiative shocks, are crucial for the formation and survival of dust.

\section{Discussion and Comparison to Observations}
\label{sec:discussion}

While novae are observed to form dust on timescales of $\sim 20-100$ days after the outburst, we find that nucleation can occur in less than 10 days after gas enters a radiative shock.  The timescale over which the majority of the ejecta will be shocked depends on the relative velocities of the dense, equatorial ejecta and the subsequent faster ejecta shell or wind, and is typically observed based on the gamma-ray emission to occur over timescales of weeks to a month (\citealt{Ackermann+14, Cheung+2016}).  Regardless of precisely where shocks form, dust formation is most likely inhibited until the gas reaches a sufficiently large distance from the WD for the temperature floor due to irradiation to decrease below the condensation temperature.  
This constraint could be alleviated if the radiation field is non-uniform$-$for instance, the nova luminosity may be preferentially directed along the polar direction instead of the dense equatorial plane where the shocks originate. In this case the gas would only be subject to the luminosity from the shock itself, allowing the gas to cool to lower temperatures and perhaps increasing the total dust yield. 
Ultimately the primary role of shocks is to create dense (low entropy), neutral regions within the ejecta, such that$-$when the appropriate temperature is reached$-$dust formation can occur efficiently.

While the high densities resulting from shock compression are favorable for forming grains, they can also be problematic insofar as they enable CO saturation, preventing carbon grain formation.  Our estimates of the efficacy of molecule destruction and our parameterization of the shock power come with large uncertainties, and this study is meant to only illustrate a range of possibilities. Indeed, dust-forming novae exhibit a variety of grain types and characteristic sizes, the range of which may be reconciled with differences in abundances and ejecta velocities.  
We expect, given our model, that a total dust event will result from a range of trajectories with variations in the composition, and the resulting grain sizes will form a distribution around some representative value(s). 
Nevertheless, even for high assumed CO destruction efficiencies $\epsilon = 10^{-2}$, we find that the amount of carbon dust formed in highly oxygen-rich ejecta is low by mass compared to that of silicate dust. 
In QV Vul (\citealt{Gehrz+92}), V705 Cas (\citealt{Mason+98}; \citealt{Evans+97}), and V1280 Sco (\citealt{Sakon+16}), carbon-rich dust is identified first, followed later by silicate formation. 
 Our models suggest that such silicate/carbon grain dichotomy in novae may be explained by chemical gradients or  heterogeneity in the ejecta (e.g.~\citealt{Gehrz+98}, \citealt{Pontefract&Rawlings04}).

When dust forms in novae, the optical depth of the grains inferred from optical extinction can range from $\lesssim 0.1$ to $\sim 10$ in the most prolific dust producing events (\citealt{McLaughlin35,Evans+05}).  Assuming that the grain size exceeds the wavelength of optical light, the cross section of a dust grain is taken to be its geometrical value of $\sigma = \pi a^{2}$.  The optical depth of a shell of grains of identical size at a radius $R$ is given by
\be
\tau = \frac{M_d\sigma }{4\pi f_{\Omega} R_{\rm sh}^{2}m_d} =\frac{3 M_d }{16\pi f_{\Omega} a \rho_d R_{\rm sh}^{2} },
\label{eq:tau}
\ee
where $m_d = (4\pi a^{3}/3)\rho_d$ is the mass of a spherical dust grain, $\rho_d \simeq 2$ g cm$^{-3}$ is the bulk density of the grain.  Inverting for the dust mass,
\be
M_d = \frac{16 \pi}{3} a \rho_d R^{2} f_{\Omega} \tau \simeq 5\times 10^{-8}M_{\odot}\left(\frac{\tau}{1}\right)\left(\frac{a}{0.1 \mu m}\right)\left(\frac{R_{\rm sh}}{3\times 10^{14}{\rm cm}}\right)^{2},
\ee  
where we have set the covering fraction $f_{\Omega} = 1/3$ and normalized the radius $R_{\rm sh}$ to a typical value of the shell radius $\sim v t$ on a timescale of a month when the dust formation is observed to occur.  Thus, we see that for $\tau \sim 0.1-10$ we require total dust masses of $M_{d} \sim 10^{-10}-10^{-7}M_{\odot}$ for 0.1$\mu$m grains.  This corresponds to a dust mass fraction in the shocked ejecta of $X_{\rm d} \sim 10^{-6}-10^{-3}$ for typical total masses of shocked gas of $M \sim 10^{-4}M_{\odot}$.  Even our most conservative model produces a total dust mass fraction exceeding $X_{\rm d} = 10^{-3}$, the majority of which is large silicate grains with sizes that exceed optical wavelengths (validating the geometric optics limit). Thus, if a fraction $\sim 0.01-1$ of the total ejecta is shocked, our model can account for the observed extinction.  

Evidence suggests that the radiative shocks in novae are concentrated in the equatorial plane, where the fast outflow meets a slow equatorial shell ($\S\ref{sec:introduction}$; \citealt{Chomiuk+14}).  Thus, even efficient dust formation may not result in strong visual extinction events for all viewers, in particular those with viewing angles aligned with the binary rotation axis.  Indeed, while only $\sim20$ percent of classical novae are classified as D-class (indicating that dust formed along the line of sight), \emph{Spitzer} observations indicate that at least 40\% of novae form dust \citep{Helton2010}. This is likely a lower limit, given that the \emph{Spitzer} sample was heavily biased towards Ne-rich novae that typically do not produce dust. Likewise, an analysis of novae in the SMARTS database (\citealt{Walter+12}) suggest that up to 80\% of non Ne-rich/recurrent novae form dust (F.~Walter, private communication). Of those, roughly $1/3$ have dust dips indicating dust on the line of sight; the other 2/3 do not (but have a near-IR excess), indicating that the dust does not form in a uniform shell.  This is consistent with a scenario in which dust formation occurs behind radiative shocks concentrated in the equatorial plane of the binary.   

 By implementing the abundances inferred from \citet{Morisset&Pequignot96}, we assume that the gas within the forward \emph{or} reverse shock originates from the WD surface. Some work suggests that the equatorial ejecta may instead be stripped from the companion \citep{Williams&Mason10}, in which case it would be primarily composed of hydrogen. However, even if the initial ejecta is lacking the necessary CNO elements, the latter ejecta, containing dredged-up material from the WD, will provide the necessary CNO products for grain formation.

Observations show that essentially no He-Ne/recurrent novae form dust.  This may be explained in our scenario by the fact that, even if internal shocks occur in these faster novae, the generally lower ejecta masses (and possibly higher ejecta velocities) of He-Ne/recurrent novae make these shocks unlikely to be radiative (eq.~\ref{eq:cool}).  Without radiative cooling, the shock-induced compression is much more modest, and the post-shock region will be less shielded from ionizing radiation.  

\citet{Sakon+16} study dust formation in the FeII nova V1280 Sco (\citealt{Munari+07}) by means of multi-epoch infrared observations.  They find that a total mass of $\sim 8\times 10^{-8}M_{\odot}$ of carbonaceous dust forms with a characteristic size of 0.01 $\mu m$, while $\sim 3\times 10^{-7}M_{\odot}$ of silicate dust forms with $0.3-0.5\mu$m.  \citet{Sakon+16} interpret the silicates as having formed in the nova ejecta, while the carbonaceous dust is either also formed in the ejecta or by the interaction between the ejecta and the circumstellar medium.  It is important to note that the observed size of the dust grains at late times may not be indicative of the their values immediately after formation due to grain erosion, e.g. due to chemical sputtering by $H^{+}$ (\citealt{Roth&GarciaRosales96}).  Still, these values are largely in agreement with those produced by our model. 

There is also suggestive direct evidence connecting shocks and dust formation in some novae.  
\citet{Weston+15} have already suggested that the presence of shocks and dust formation may be related in the case of nova V1723 Aql, which showed evidence for shocks via x-ray and radio synchrotron emission at $t\sim40\pm20$ days after outburst and formed dust around $t=20$ days (\citealt{Nagashima+2013}).
The most luminous gamma-ray nova to date, V1324 Sco, began emitting gamma rays at $t\sim15$ days after the outburst. At $t\sim45$ days, it displayed a significant dust extinction event that decreased the optical and near-infrared flux by $\sim8.5$ magnitudes (\citealt{Finzell+2017}).  
The `cuspy' and peculiar nova V2362 Cyg (\citealt{Lynch+08}) formed dust during the second peak in its optical light curve, at the same time that hard X-rays were observed indicating the presence of shocks.  V5668 Sgr also showed hard X-rays immediately prior to an extinction event producing $\sim3\times10^{-7} M_{\odot}$ of dust (\citealt{Banerjee16}).  \citet{Evans+2017} study the infrared emission from V339 Del, finding dust formation beginning at $t\sim35$ days after the gamma ray emission peaked at $t\sim6$ days after optical maximum. No X-ray emission was detected by \emph{Swift} until 30 days after the eruption, at which time it was characteristic with shocked gas from an expanding shell.

Dust formation in post-shock gas of radiative shocks is not a phenomenon exclusive to novae.  Observations of rapid dust formation coincident with shock-powered X-ray emission exist for some supernovae that show strong circumstellar interaction (e.g. 2005ip in \citealt{Fox09} and 2006jc in \citealt{Smith+08}). In some cases this is also accompanied by episodic He II ($4686 \AA$) emission that can be associated with the radiative cooling in the post-shock gas. This is also observed in colliding stellar winds, particularly WC+O binaries {\citep{Smith10}.  

\section{Conclusions}
\label{sec:conclusions}

We have explored radiative shocks in classical novae as environments for dust formation.  Our results are summarized as follows.
\begin{itemize}
\item Evidence from across the electromagnetic spectrum, most recently gamma-rays from {\it Fermi} LAT, suggests that internal shocks are common, if not ubiquitous, features of classical novae.  The ejecta is sufficiently dense at the time of the observed gamma-ray emission that these shocks are likely to be radiative.  A substantial fraction of the nova ejecta must pass through such a shock in order to explain the high observed gamma-ray luminosities.  
\item As gas behind the radiative shock cools, it compresses, reaching densities up to $n\sim10^{14}\rm cm^{-3}$.  The shocked ejecta collects in a geometrically-thin, clumpy shell.  Though commonly inferred observationally (e.g.~\citealt{Friedjung+99}), such thin-shell geometry is challenging to understand if the ejecta were to expand freely from the WD surface.  
\item We develop a simple one-zone model for the thermodynamic evolution of typical fluid elements after it passes through the shock, accounting for radiative cooling and irradiation from the central white dwarf.  Applying a simple molecular chemistry network and classical nucleation theory to these  density/temperature trajectories, we show that dust formation can occur efficiently within the post-shock gas.
\item Given the very high densities of the post-shock gas, dust formation is regulated by CO chemistry, with oxygen-rich regions forming primarily silicate grains.  Carbon grain formation is inhibited by CO formation, unless the C:O ratio is $\gtrsim 1$.  After nucleation commences, grains easily grow to sizes of up to $a \sim 0.1\mu$m or greater, given the appropriate abundances.  The total dust mass formed is sufficient to explain the IR emission from novae and the surface area is sufficient to explain even large extinction events.
\item High-energy particles accelerated at the shock can destroy CO, allowing for the formation of both grain types, even in carbon-poor gas.  However, we estimate that the effects of non-thermal destruction, though uncertain, are relatively weak and hence play a less prominent role in carbonaceous/silicate dust dichotomy in novae than in supernovae (e.g.~\citealt{Lazzati&Heger16}).
\item UV and X-ray radiation from the central WD is unlikely to penetrate and ionize the ejecta until the density decreases to $n \lesssim 10^{8}-10^{9}$ cm$^{-3}$ (eq.~\ref{eq:nion}), after which time the temperature is already well below the threshold for carbonaceous or silicate dust formation and the ejecta has expanded to very large radii.  
\item Radiative shocks are more likely to occur in novae with higher ejecta masses and lower ejecta velocities since line cooling increases rapidly for lower shock velocities.  This could help explain the prevalence of dust formation in slower novae (from CO WDs) and its apparent absence in faster He/Ne or recurrent novae.  If internal shocks are concentrated in the equatorial plane of the binary, then visual extinction events may preferentially occur for viewing angles within this plane.
\end{itemize}

\section*{Acknowledgments}
AMD acknowledges support by NSF Graduate Research Fellowship under Grant Number DGE-16-44869.  BDM gratefully acknowledges support from NASA grants NNX15AU77G (Fermi), NNX15AR47G (Swift), and NNX16AB30G (ATP), NSF grant AST-1410950, and the Alfred P. Sloan Foundation.  We thank Kengo Tomida and Ondrej Pejcha for supplying the tabulated opacities and EOS used in our calculations.  We thank Thomas Finzell, Jeno Sokoloski, Andrew Helton, John Raymond, Fred Walter, and Indrek Vurm for helpful information and inspiring conversations.

\appendix

\section{Shock-Generated Electrons in Molecular Zone}
\label{sec:electrons}

In this section, we estimate the fraction of the shock kinetic power which is placed into relativistic particles and is available to dissociate CO molecules.

In the standard theory of Diffusive Shock Acceleration (DSA), particles are accelerated to energies exceeding that of the thermal plasma by diffusing back and forth across the shock front via interaction with magnetic turbulence upstream and downstream.  The momentum distribution, $f(p) \propto p^{-4}$, predicted by the simplest DSA models corresponds to an energy distribution
\begin{eqnarray}
&& \frac{dN}{dE}E^{2} \propto  
\left\{
\begin{array}{lr}
E^{1/2}
 &
kT_{\rm sh} \lesssim E \ll m c^{2} \\
{\rm constant,} &
m c^{2} \ll E < E_{\rm max}, \\
\end{array}
\right..
\label{eq:dNdEion}
\end{eqnarray}
which concentrates most of the non-thermal energy in relativistic particles of energy $E \gtrsim m c^{2}$, where $m$ is the mass of the electron or ion and $E_{\rm max} \sim 10^{10}-10^{12}$ eV is the maximum energy to which particles are accelerated (\citealt{Metzger+16}).  

Relativistic leptons can originate either directly from the acceleration of electrons at the shock, or indirectly from $e^{\pm}$ pairs which are produced by the decay of charged pions from inelastic proton-proton collisions.  Modeling of the nova gamma-ray  (\citealt{Ackermann+14}; \citealt{Metzger+15}) and radio (\citealt{Vlasov+2016}) emission is consistent with a fraction $\epsilon_{\rm e} \sim 10^{-2}$ and $\epsilon_{\rm p} \sim 0.1$ of the shock power being placed into non-thermal electrons and protons, respectively.  The fraction of the shock power given to $e^{\pm}$ pairs is thus $\epsilon_{\pm} \simeq f_{\pi}\epsilon_p/3$, where $f_{\pi} \approx 0.1$ is the fraction of the proton energy per inelastic collision placed into pions, and the factor of $1/3$ accounts for the the other $2/3$ of the pion energy, which goes into photons ($\pi_0$) and neutrinos. 

First, consider relativistic electrons that are directly accelerated at the shock and then advected downstream with the flow.  An electron produced near the shock of initial energy $E_e$ will lose that energy to inelastic Coulomb scatterings with the thermal plasma on a timescale given by 
\be
t_{\rm C} \approx \left\{
\begin{array}{lr}
150(E_e/m_e c^{2})n_{10}^{-1}\,{\rm s}
, &
E_e \gg m_e c^{2}\\
100(E_e/m_e c^{2})^{3/2}n_{10}^{-1}\,\,{\rm s}, &
E_e \ll m_e c^{2}, \\
\end{array}
\label{eq:tC}
\right..
\ee
where $n_e \approx n = 10^{10}n_{10}$ cm$^{-3}$ is the post-shock electron density and we have adopted a value of ln $\Lambda \approx 25$ for the Coulomb logarithm.  This timescale is generally shorter than the cooling timescale over which the post-shock gas cools and enters the central shell, $t_{\rm cool} \approx 8000n_{10}^{-1} v_8^{3.4}$ s (eq.~\ref{eq:tcool}, for $v_8 < 1$), except for particles of energy\footnote{At much higher energies, $E_e \gtrsim 10^{3}m_e c^{2}$, additional cooling from relativistic bremsstrahlung emission dominates Coulomb cooling in the compressing gas (\citealt{VurmMetzger2016}).}
\be
E_e \gtrsim E_{\rm min,lep} \approx 53m_e c^{2}v_8^{3.4}.
\label{eq:emaxlep}
\ee
Depending on the shock speed, electrons will enter the cool molecular shell with energies $E_e \gtrsim 1-100 m_e c^{2}$.

Now consider $e^{\pm}$ pairs which are produced indirectly through hadronic interactions.  Relativistic protons accelerated at the shock to collide with background protons of density $\sim n$ on a timescale of
\be t_{\pi} = (n \sigma_{\pi}c)^{-1} = 8\times 10^{5}n_{10}^{-1}\,{\rm s},
\ee
where $\sigma_{\pi } \approx 4\times 10^{-26}$ cm$^{2}$ is the inelastic p-p cross section (\citealt{Kamae+06}).  The value of $t_{\pi}$ is much longer than the flow time behind the shock $\sim t_{\rm cool}$, indicating that pion production will mainly occur within the central shell (\citealt{Metzger+16}).  Charged pions decay into pairs of energy $\gtrsim 0.3 m_{\pi} c^{2} \sim 40$ MeV (e.g., \citealt{Vlasov+2016}), such that, in analogy with equation (\ref{eq:emaxlep}), the minimum energy of pairs which are injected into the molecular zone is given by
\be
E_e \gtrsim E_{\rm min, had} \approx 0.3 m_{\pi} c^{2} \approx 40 {\rm MeV}
\ee 

Once in the molecular region, relativistic electrons or positrons of energy $E_e \gtrsim E_{\rm min, lep}, E_{\rm min,had}$ lose energy to molecular dissociation and ionization in addition to Coulomb energy losses (eq.~\ref{eq:tC}).  The timescale for energy loss to dissociation is given by
\be
t_{\rm dis} \approx \frac{E_e}{L n_{\rm mol} v_{\rm e}}  \approx
\left\{
\begin{array}{lr}
 200(E_{e}/m_e c^{2})^{2}n_{\rm mol,10}^{-1}\,{\rm s}, &
E_e \gg m_e c^{2} \\
150(E_{e}/m_e c^{2})^{3/2}n_{\rm mol,10}^{-1}\,{\rm s}, &
E_e \ll m_e c^{2}, \\
\end{array}
\right..
\ee
where $n_{\rm mol}$ is the density of molecules and $L \approx 4\times 10^{-15}(E_e/{\rm keV})^{-1}$ cm$^{2}$ eV is the energy loss function, which we have obtained by extrapolating Fig.~3 of Victor \& Liu (1994) to high energies $E_e \gtrsim$ keV.

While the electrons or positrons are still ultra-relativistic ($E_e \gg m_e c^{2}$), we have $t_{\rm C}/t_{\rm dis} \propto (E/m_e c^{2})^{-1} \ll 1$, and hence Coulomb losses dominate ionization losses.  However, once the electrons cool to mildly relativistic energies ($E_e \lesssim m_e c^{2}$) a fixed fraction $t_{\rm C}/t_{\rm dis} \approx n_{\rm mol}/(n + n_{\rm mol})$ of their power goes into dissociation/ionization.  Thus, the total fraction of the initial electron/positron energy used to dissociate molecules is very roughly approximated as
\be
f_{\rm dis} \sim \left(\frac{E_{\rm min}}{m_e c^{2}}\right)^{-1} f_{\rm mol},
\ee
where $E_{\rm min} = E_{\rm min,lep}$ or $E_{\rm min,had}$ and $f_{\rm mol} \equiv \frac{n_{\rm mol}}{(n + n_{\rm mol})}$ is the number fraction of the relevant molecule.

Combining the fact that electrons(positrons) of energy $\gtrsim E_{\rm min}$ represent an order-unity fraction of the total shock-accelerated population for the standard DSA spectrum (eq.~\ref{eq:dNdEion}), we conclude that, to order of magnitude, the fraction of the shock power which goes into dissociating molecules is given in the leptonic case by
\be
\epsilon \approx \epsilon_e 
f_{\rm dis} \sim \left(\frac{E_{\rm min,lep}}{m_e c^{2}}\right)^{-1}f_{\rm mol} \sim 10^{-3} v_8^{-3.4}f_{\rm mol},
\ee
In the hadronic case, the fraction is 
\be
\epsilon \approx \left(\frac{\epsilon_p}{3}\right)f_{\pi}\left(\frac{E_{\rm min,had}}{m_e c^{2}}\right)^{-1}f_{\rm mol} \sim 10^{-4}f_{\rm mol},
\ee
where again we have averaged over the pair distribution produced by a logarithmically flat proton energy distribution (eq.~\ref{eq:dNdEion}).  

Finally, protons accelerated at the shock could themselves directly dissociate molecules in the central shell.  For high energy $\gtrsim m_\pi c^{2}$ protons, the pion production cross section is much higher than the molecular cross section.  However, for lower energy protons of energy $\ll m_{\pi} c^{2}$ below the pion production threshold, perhaps constituting one percent of the total energy placed in non-thermal protons, a significant fraction of the proton energy could be available to ionize atoms or molecules.  We therefore crudely estimate a CO dissociation efficiency of
\be
\epsilon \lesssim 0.01\epsilon_p f_{\rm mol} \sim 3\times 10^{-3} f_{\rm mol},
\ee
comparable to leptons produced in the direct acceleration case.

For a typical CO molecule fraction of $f_{\rm mol} \sim 0.1$, we estimate total characteristic values of $\epsilon \sim 10^{-4}-3\times 10^{-3}$.  These motivate the fiducial value of $\epsilon = 10^{-4}$ adopted in our chemistry network (Section \ref{sec:chemical}).   We assume in equation (\ref{eq:Knth}) that the dissociation rate directly tracks the shock power.  However, the {\it effective} value of $\epsilon$ could be higher than estimated above if, for instance, the pion production timescale within the central shell is longer than the evolution timescale over which the shock power is dropping.  In such a case, the rate of dissociation could be potentially higher than would be predicted by the {\it current} shock power.

\section{Ionization State}
\label{sec:ionization}

Here we address the density above which the ejecta will remain opaque to hard radiation above the ionization thresholds of abundant elements such as hydrogen, helium, and carbon.  The latter being neutral is thought to be a key requisite to the chemistry required to form complex molecules and dust.  As the density of the ejecta decreases due to radial expansion, ionizing super-soft X-ray radiation from the central white dwarf  (e.g.~\citealt{Schwarz+11}, \citealt{Wolf+13}) will eventually penetrate this material.  

Absent an external source of photo-ionization, atoms will recombine into their ground states in the central shell.  The optical depth through the freely expanding central shell at time $t$ is given by
\be
\tau = \frac{M_{\rm ej}\kappa}{4\pi R_{\rm sh}^{2}} \approx 0.23 \left(\frac{\kappa}{\rm cm^{2}\,g^{-1}}\right) M_{-4}v_{8}^{-2}t_{\rm month}^{-2} ,
\ee
where $\kappa$ is the opacity, $R_{\rm sh} = v_{\rm sh}t$, $v_{\rm sh} = 10^{8}v_8$ cm s$^{-1}$ and $t = t_{\rm month}$ month.

For hydrogen-like species of charge $Z$, atomic mass $A$ and mass fraction $X_A$, the bound-free opacity of neutral gas is approximately given by  (\citealt{Osterbrock&Ferland06})
\be \kappa_{\rm bf} \approx \frac{X_{A}\sigma_{\rm bf,\nu}f_{\rm n}}{Am_p} \approx 3.6\times 10^{6}\,{\rm cm^{2}\,g^{-1}}\,f_{\rm n}X_A A^{-1}Z^{-2}\left(\frac{\nu_{\rm thr}}{\nu}\right)^{3},
\label{eq:sigmabf}
\ee
where $\sigma_{\rm bf,\nu}$ is the bound-free cross section for photons of frequency $\nu$, $\nu_{\rm thr} \approx 13.6Z^{2}$ eV is the ionization threshold frequency and $f_{\rm n}$ is the neutral fraction.  
For values of the opacity $\kappa \approx \kappa_{\rm bf}$ corresponding to neutral ejecta with $f_n \sim 1$, we see that the shell will remain optically thick $\tau \gg 1$ to UV/X-ray photons for many months.  

Hard radiation can penetrate the ejecta once the neutral fraction is reduced to a value $f_{\rm n} \ll 1$ by photo-ionization.  The importance of photo-ionization is commonly quantified by the ionization parameter, which for the cold central shell is given by
\begin{eqnarray}
 \xi = \frac{4\pi F_{\rm H}}{n} = 0.14\,{\rm erg\, cm \,s^{-1}}\,\,L_{38}n_{10}^{-1}v_{8}^{-2}t_{\rm month}^{-2}
\label{eq:xi}
\end{eqnarray} 
where $F_{\rm H} \approx L_{\rm WD}/4\pi R_{\rm sh}^{2}$ is the ionizing flux from the WD of luminosity $L_{\rm WD} = 10^{38}L_{38}$ erg s$^{-1}$.  An ionization parameter $\xi \gg 1$ is required for complete ionization, depending on composition.

Nebular radiation will ionize the ejecta to a depth $\Delta$, set by the location at which the effective optical depth of a photon of frequency $\nu \gtrsim \nu_{\rm thr}$ to absorption reaches unity, i.e.,
\be
1 = \int_{0}^{\Delta}\rho_{\rm ej}\kappa_{\rm bf,\nu}\left[1 +\rho_{\rm ej}\kappa_{\rm es}s \right]ds \approx \tau_{\rm abs}(1 + \tau_{\rm es}),
\label{eq:tau1}
\ee
where $s$ is the depth through the ionized layer, $\tau_{\rm abs} \equiv \kappa_{\rm bf,\nu}\rho_{\rm ej}\Delta$ is the optical depth through the layer to true absorption, and $\rho_{\rm ej} = \mu n m_p$ is the density in the shell (assumed uniform).  The factor $1 + \tau_{\rm es}$ accounts for the additional path-length traversed by the photon due to electron scattering, where
$\tau_{\rm es} = \rho_{\rm ej}\kappa_{\rm es}\Delta$ is the electron scattering optical depth through the ionized layer.    

The neutral fraction $f_n$ is determined by the balance between photo-ionization and recombination:
\begin{eqnarray}
f_n &=& \left(1 + \frac{4\pi}{\alpha_{\rm rec} n_e}\int\frac{J_{\nu}}{h\nu}\sigma_{\rm bf,\nu}d\nu\right)^{-1} \nonumber \\ &\underset{f_{n} \ll 1}\approx& \alpha_{\rm rec} n_e \pi R_{\rm sh}^{2} \left(\int_{\nu_{\rm thr}}^{\infty}\frac{L_{\nu}}{h\nu}\sigma_{\rm bf,\nu}d\nu\right)^{-1}  \approx \frac{\pi R_{\rm sh}^{2}\alpha_{\rm rec} n_e(h \nu_{\rm th})}{L_{\nu_{\rm th}}\nu_{\rm th}\sigma_{\rm bf,th}} \nonumber \\ &\approx& 2\times 10^{-5}Z^{6}\epsilon_{\rm thr}^{-1} T_{e,4}^{-0.8}n_{10}v_{8}^{2}t_{\rm month}^{2}L_{38}^{-1}
\label{eq:fn}
\end{eqnarray}
where $J_{\nu} = L_{\nu}/4\pi^{2} R_{\rm sh}^{2}$ is the mean specific intensity of ionizing radiation, $\alpha_{\rm rec} = 2.6\times 10^{-13}Z^{2}T_{e,4}^{-0.8}$ cm$^{3}$ s$^{-1}$ is the approximate case B radiative recombination rate, and $T_{e,4}$ is the electron temperature in the ionized layer in units of $10^{4}$ K.  We define $\epsilon_{\rm thr} \equiv  \nu_{\rm thr}L_{\nu_{\rm thr}}/L_{\rm WD}< 1$ as the fraction of the total ionizing radiation which resides at frequencies near the ionization threshold.

Using equations (\ref{eq:sigmabf}) and (\ref{eq:fn}), the ratio of bound-free to scattering opacity at the ionization threshold is given by
\begin{eqnarray}
&\zeta_{\rm thr} &\equiv \left.\frac{\kappa_{\rm bf,\nu}}{\kappa_{\rm es}}\right|_{\nu = \nu_{\rm thr}} \simeq 
\frac{X_{A}\sigma_{\rm bf,\nu_{\rm thr}}f_{\rm n}}{A m_p \kappa_{\rm es}} \nonumber \\
&\approx&  180 X_{A}A^{-1}Z^{4}\epsilon_{\rm thr}^{-1} T_{e,4}^{-0.8}n_{10}v_{8}^{2}t_{\rm month}^{2}L_{X,38}^{-1} \gg 1,
\label{eq:zeta}
\end{eqnarray}
where we have estimated the number density of electrons as $n_e \sim n$.  Bound-free opacity thus dominates electron scattering opacity for parameters of interest, allowing us to neglect the effect of the latter on the opacity depth in equation (\ref{eq:tau1}).  

Photons of frequency $\nu_{\rm th}$ penetrate the ejecta to a depth $\Delta_{\rm ion} \approx \rho_{\rm ej}/\kappa_{\rm bf}$, the ratio with respect to the ejecta radius is given by
\begin{eqnarray}
\frac{\Delta_{\rm ion}}{\Delta_{\rm sh}} &\approx& \frac{4\pi R_{\rm sh}^{2}}{M_{\rm ej}\kappa_{\rm bf}} \approx \frac{L_{\nu_{\rm th}}\nu_{\rm th}A m_p}{M_{\rm ej}X_A \alpha_{\rm rec}n_e (h \nu_{\rm th})} \nonumber \\
& \approx& 0.06 A X_{A}^{-1}Z^{-4}\epsilon_{\rm thr} M_{-4}^{-1}T_{e,4}^{0.8}n_{10}^{-1}L_{38},
\end{eqnarray}
where $\Delta_{\rm sh} = M_{\rm ej}/(4\pi \rho_{\rm ej} R_{\rm sh}^{2})$ is the shell thickness.  The shell will shield ionizing radiation from its interior until $\Delta_{\rm ion} \sim \Delta_{\rm sh}$, as occurs below a critical density $n_{\rm ion}$ given by
\begin{eqnarray}
n_{\rm ion} &\approx& \frac{\epsilon_{\rm thr}L_{\rm WD}A m_p}{(h\nu_{\rm th})\alpha_{\rm rec}M_{\rm ej}X_A} \nonumber \\
&\approx& 6\times 10^{8}{\rm cm^{-3}}M_{-4}^{-1}\left(\frac{\epsilon_{\rm thr}}{0.1}\right)L_{38} Z^{-4} A \left(\frac{X_{A}}{0.1}\right)^{-1}T_{e,4}^{0.8} \nonumber \\
\end{eqnarray}
Now, adopt a normalization appropriate to CI, for which $h\nu_{\rm thr} \approx 11.3$ eV, the threshold cross section $\sigma_{\rm th} \approx$ 10Mb, and the recombination rate is $6.9\times 10^{-13}$ cm$^{3}$ s$^{-1}$ for $\approx 10^{4}$ K gas \citep{Nahar96}.  Adopting these values, we find that a carbon ionization front penetrates the ejecta below a density of
\be
n_{\rm ion, C} \approx  8\times 10^{8}{\rm cm^{-3}}M_{-4}^{-1}(\epsilon_{\rm thr}/0.1)L_{38} (X_{C}/0.1)^{-1}
\label{eq:nion}
\ee
Thus, the ejecta of mass $\sim 10^{-5}-10^{-4}M_{\odot}$ will remain shielded from H-ionizing 13.6 eV photons ($Z = 1$, $A = 1$, $X_A \sim 1$) until the $n \lesssim 10^{9}-10^{10}$ cm$^{-3}$ and shielded from carbon ionizing photons to a similar density range.  As shown in Figure \ref{fig:thermo}, these densities are only achieved at sufficiently large radii where dust has already formed.

It is useful to compare the minimum shielding density to the ejecta density at the time of dust condensation (see Ch.~13 by Evans \& Rawlings in \citealt{Bode&Evans08}). The temperature $T_d$ of a dust grain in the central shell a distance $v_{\rm ej}t$ from the star is given by
\be
T_{\rm d} = \left(\frac{L}{16\pi r^{2}\sigma}\frac{\langle Q_a \rangle}{\langle Q_e \rangle}\right)^{1/4},
\ee
where $\langle Q_a \rangle$ and $\langle Q_e \rangle$ are the Planck mean absorptivity and emissivity of the grain material, respectively.  Taking $\langle Q_a \rangle = 1$ and $\langle Q_e \rangle \approx 0.01 a T^{2}$ gives a condensation radius ($T_d \approx T_c $) of
\begin{eqnarray}
R_{\rm c} &=& \left(\frac{L }{16\pi \sigma T_{\rm c}^{4}}\frac{\langle Q_a \rangle}{\langle Q_e \rangle}\right)^{1/2}  \nonumber \\
&\approx& 1.0\times 10^{14}{\rm cm}\,\,L_{38}^{1/2}\left(\frac{T_{c}}{1200{\rm K}}\right)^{-3}\left(\frac{b}{1\mu {\rm m}}\right)^{-1/2},
\end{eqnarray}
where the condensation temperature $T_c = 1200$ K is normalized to a characteristic value for graphite.  

The shell reaches this radius on the expansion timescale of
\be
t_{\rm c} = \frac{R_{\rm c}}{v_{\rm ej}} \approx 12{\rm d}\,\,v_8 L_{38}^{1/2}\left(\frac{T_{c}}{1200{\rm K}}\right)^{-3}\left(\frac{a}{1\mu {\rm m}}\right)^{-1/2},
\ee
at which time the mean density of the (unshocked) nova ejecta is given by
\be
n_{\rm c} \approx \frac{3 M_{\rm ej}}{4\pi R_{\rm c}^{3} m_p} \approx 3\times 10^{10}\,{\rm cm^{-3}},M_{-4}L_{38}^{-3/2}\left(\frac{T_{c}}{1200{\rm K}}\right)^{9}\left(\frac{a}{1\mu {\rm m}}\right)^{3/2}
\ee
Thus, for small grains $a \lesssim 0.1\mu$m we have $n_{c} \lesssim n_{\rm ion}$, {\it indicating that such grains cannot form given the mean densities we expect if the nova ejecta expands as a uniform spherical outflow}.  This emphasizes the need for radiative shocks, which compress the gas by factors of $\gtrsim 10^{3}$ (Section \ref{sec:overdensities}).

\bibliographystyle{mnras}

\bibliography{novabib}

\begin{thebibliography}{}
\makeatletter
\relax
\def\mn@urlcharsother{\let\do\@makeother \do\$\do\&\do\#\do\^\do\_\do\%\do\~}
\def\mn@doi{\begingroup\mn@urlcharsother \@ifnextchar [ {\mn@doi@}
  {\mn@doi@[]}}
\def\mn@doi@[#1]#2{\def\@tempa{#1}\ifx\@tempa\@empty \href
  {http://dx.doi.org/#2} {doi:#2}\else \href {http://dx.doi.org/#2} {#1}\fi
  \endgroup}
\def\mn@eprint#1#2{\mn@eprint@#1:#2::\@nil}
\def\mn@eprint@arXiv#1{\href {http://arxiv.org/abs/#1} {{\tt arXiv:#1}}}
\def\mn@eprint@dblp#1{\href {http://dblp.uni-trier.de/rec/bibtex/#1.xml}
  {dblp:#1}}
\def\mn@eprint@#1:#2:#3:#4\@nil{\def\@tempa {#1}\def\@tempb {#2}\def\@tempc
  {#3}\ifx \@tempc \@empty \let \@tempc \@tempb \let \@tempb \@tempa \fi \ifx
  \@tempb \@empty \def\@tempb {arXiv}\fi \@ifundefined
  {mn@eprint@\@tempb}{\@tempb:\@tempc}{\expandafter \expandafter \csname
  mn@eprint@\@tempb\endcsname \expandafter{\@tempc}}}

\bibitem[\protect\citeauthoryear{{Abdo} et~al.,}{{Abdo} et~al.}{2010}]{Abdo+10}
{Abdo} A.~A.,  et~al., 2010, \mn@doi [Science] {10.1126/science.1192537}, \href
  {http://adsabs.harvard.edu/abs/2010Sci...329..817A} {329, 817}

\bibitem[\protect\citeauthoryear{{Ackermann} et~al.}{{Ackermann}
  et~al.}{2014}]{Ackermann+14}
{Ackermann} M.,  et~al., 2014, \mn@doi [Science] {10.1126/science.1253947},
  \href {http://adsabs.harvard.edu/abs/2014Sci...345..554A} {345, 554}

\bibitem[\protect\citeauthoryear{{Banerjee}, {Srivastava}, {Ashok}  \&
  {Venkataraman}}{{Banerjee} et~al.}{2016}]{Banerjee16}
{Banerjee} D.~P.~K.,  {Srivastava} M.~K.,  {Ashok} N.~M.,   {Venkataraman} V.,
  2016, \mn@doi [\mnras] {10.1093/mnrasl/slv163}, \href
  {http://adsabs.harvard.edu/abs/2016MNRAS.455L.109B} {455, L109}

\bibitem[\protect\citeauthoryear{{Bath} \& {Harkness}}{{Bath} \&
  {Harkness}}{1989}]{Bath&Harkness89}
{Bath} G.~T.,  {Harkness} R.~P.,  1989, in {Bode} M.~F.,  {Evans} A.,  eds,
  Classical Novae. pp 61--72

\bibitem[\protect\citeauthoryear{{Bath} \& {Shaviv}}{{Bath} \&
  {Shaviv}}{1976}]{Bath&Shaviv76}
{Bath} G.~T.,  {Shaviv} G.,  1976, \mnras, \href
  {http://adsabs.harvard.edu/abs/1976MNRAS.175..305B} {175, 305}

\bibitem[\protect\citeauthoryear{{Becker} \& {D{\"o}ring}}{{Becker} \&
  {D{\"o}ring}}{1935}]{BeckerDoring35}
{Becker} R.,  {D{\"o}ring} W.,  1935, Annalen der Physik, 24, 719

\bibitem[\protect\citeauthoryear{{Bode} \& {Evans}}{{Bode} \&
  {Evans}}{2008}]{Bode&Evans08}
{Bode} M.~F.,  {Evans} A.,  2008, {Classical Novae}

\bibitem[\protect\citeauthoryear{{Bode}, {Roberts}, {Whittet}, {Seaquist}  \&
  {Frail}}{{Bode} et~al.}{1987}]{Bode+87}
{Bode} M.~F.,  {Roberts} J.~A.,  {Whittet} D.~C.~B.,  {Seaquist} E.~R.,
  {Frail} D.~A.,  1987, \mn@doi [\nat] {10.1038/329519a0}, \href
  {http://adsabs.harvard.edu/abs/1987Natur.329..519B} {329, 519}

\bibitem[\protect\citeauthoryear{{Cherchneff} \& {Dwek}}{{Cherchneff} \&
  {Dwek}}{2009}]{Cherchneff&Dwek2009}
{Cherchneff} I.,  {Dwek} E.,  2009, \mn@doi [\apj]
  {10.1088/0004-637X/703/1/642}, \href
  {http://adsabs.harvard.edu/abs/2009ApJ...703..642C} {703, 642}

\bibitem[\protect\citeauthoryear{{Cheung} et~al.,}{{Cheung}
  et~al.}{2016}]{Cheung+2016}
{Cheung} C.~C.,  et~al., 2016, preprint, \href
  {http://adsabs.harvard.edu/abs/2016arXiv160504216C} {} (\mn@eprint {arXiv}
  {1605.04216})

\bibitem[\protect\citeauthoryear{{Chevalier} \& {Imamura}}{{Chevalier} \&
  {Imamura}}{1982}]{Chevalier&Imamura82}
{Chevalier} R.~A.,  {Imamura} J.~N.,  1982, \mn@doi [\apj] {10.1086/160364},
  \href {http://adsabs.harvard.edu/abs/1982ApJ...261..543C} {261, 543}

\bibitem[\protect\citeauthoryear{{Chomiuk} et~al.,}{{Chomiuk}
  et~al.}{2014}]{Chomiuk+14}
{Chomiuk} L.,  et~al., 2014, \mn@doi [\nat] {10.1038/nature13773}, \href
  {http://adsabs.harvard.edu/abs/2014Natur.514..339C} {514, 339}

\bibitem[\protect\citeauthoryear{{Clayton}}{{Clayton}}{2013}]{Clayton2013}
{Clayton} D.~D.,  2013, \mn@doi [\apj] {10.1088/0004-637X/762/1/5}, \href
  {http://adsabs.harvard.edu/abs/2013ApJ...762....5C} {762, 5}

\bibitem[\protect\citeauthoryear{{Donn} \& {Nuth}}{{Donn} \&
  {Nuth}}{1985}]{DonnNuth1985}
{Donn} B.,  {Nuth} J.~A.,  1985, \mn@doi [\apj] {10.1086/162779}, \href
  {http://adsabs.harvard.edu/abs/1985ApJ...288..187D} {288, 187}

\bibitem[\protect\citeauthoryear{{Dougherty}, {Waters}, {Bode}, {Lloyd},
  {Kester}  \& {Bontekoe}}{{Dougherty} et~al.}{1996}]{Dougherty+96}
{Dougherty} S.~M.,  {Waters} L.~B.~F.~M.,  {Bode} M.~F.,  {Lloyd} H.~M.,
  {Kester} D.~J.~M.,   {Bontekoe} T.~R.,  1996, \aap, \href
  {http://adsabs.harvard.edu/abs/1996A%26A...306..547D} {306, 547}

\bibitem[\protect\citeauthoryear{{Drake} et~al.,}{{Drake}
  et~al.}{2009}]{Drake+09}
{Drake} J.~J.,  et~al., 2009, \mn@doi [\apj] {10.1088/0004-637X/691/1/418},
  \href {http://adsabs.harvard.edu/abs/2009ApJ...691..418D} {691, 418}

\bibitem[\protect\citeauthoryear{{Evans} \& {Gehrz}}{{Evans} \&
  {Gehrz}}{2008}]{Evans&Gehrz08}
{Evans} A.,  {Gehrz} R.~D.,  2008, {in Classical Novae}

\bibitem[\protect\citeauthoryear{{Evans} \& {Gehrz}}{{Evans} \&
  {Gehrz}}{2012}]{Evans&Gehrz12}
{Evans} A.,  {Gehrz} R.~D.,  2012, Bulletin of the Astronomical Society of
  India, \href {http://adsabs.harvard.edu/abs/2012BASI...40..213E} {40, 213}

\bibitem[\protect\citeauthoryear{{Evans}, {Geballe}, {Rawlings}  \&
  {Scott}}{{Evans} et~al.}{1996}]{Evans+96}
{Evans} A.,  {Geballe} T.~R.,  {Rawlings} J.~M.~C.,   {Scott} A.~D.,  1996,
  \mn@doi [\mnras] {10.1093/mnras/282.3.1049}, \href
  {http://adsabs.harvard.edu/abs/1996MNRAS.282.1049E} {282, 1049}

\bibitem[\protect\citeauthoryear{{Evans}, {Geballe}, {Rawlings}, {Eyres}  \&
  {Davies}}{{Evans} et~al.}{1997}]{Evans+97}
{Evans} A.,  {Geballe} T.~R.,  {Rawlings} J.~M.~C.,  {Eyres} S.~P.~S.,
  {Davies} J.~K.,  1997, \mn@doi [\mnras] {10.1093/mnras/292.1.192}, \href
  {http://adsabs.harvard.edu/abs/1997MNRAS.292..192E} {292, 192}

\bibitem[\protect\citeauthoryear{{Evans}, {Tyne}, {Smith}, {Geballe},
  {Rawlings}  \& {Eyres}}{{Evans} et~al.}{2005}]{Evans+05}
{Evans} A.,  {Tyne} V.~H.,  {Smith} O.,  {Geballe} T.~R.,  {Rawlings} J.~M.~C.,
    {Eyres} S.~P.~S.,  2005, \mn@doi [\mnras]
  {10.1111/j.1365-2966.2005.09146.x}, \href
  {http://adsabs.harvard.edu/abs/2005MNRAS.360.1483E} {360, 1483}

\bibitem[\protect\citeauthoryear{{Evans} et~al.,}{{Evans}
  et~al.}{2017}]{Evans+2017}
{Evans} A.,  et~al., 2017, \mn@doi [\mnras] {10.1093/mnras/stw3334}, \href
  {http://adsabs.harvard.edu/abs/2017MNRAS.tmp...67E} {}

\bibitem[\protect\citeauthoryear{Feder, Russell, Lothe  \& Pound}{Feder
  et~al.}{1966}]{Feder66}
Feder J.,  Russell K.,  Lothe J.,   Pound G.,  1966, \mn@doi [Advances in
  Physics] {10.1080/00018736600101264}, 15, 111

\bibitem[\protect\citeauthoryear{{Ferguson}, {Alexander}, {Allard}, {Barman},
  {Bodnarik}, {Hauschildt}, {Heffner-Wong}  \& {Tamanai}}{{Ferguson}
  et~al.}{2005}]{Ferguson+05}
{Ferguson} J.~W.,  {Alexander} D.~R.,  {Allard} F.,  {Barman} T.,  {Bodnarik}
  J.~G.,  {Hauschildt} P.~H.,  {Heffner-Wong} A.,   {Tamanai} A.,  2005,
  \mn@doi [\apj] {10.1086/428642}, \href
  {http://adsabs.harvard.edu/abs/2005ApJ...623..585F} {623, 585}

\bibitem[\protect\citeauthoryear{{Ferland} \& {Shields}}{{Ferland} \&
  {Shields}}{1978}]{Ferland&Shields78}
{Ferland} G.~J.,  {Shields} G.~A.,  1978, \mn@doi [\apj] {10.1086/156597},
  \href {http://adsabs.harvard.edu/abs/1978ApJ...226..172F} {226, 172}

\bibitem[\protect\citeauthoryear{{Finzell}, {Chomiuk}, {Munari}  \&
  {Walter}}{{Finzell} et~al.}{2015}]{Finzell+15}
{Finzell} T.,  {Chomiuk} L.,  {Munari} U.,   {Walter} F.~M.,  2015, \mn@doi
  [\apj] {10.1088/0004-637X/809/2/160}, \href
  {http://adsabs.harvard.edu/abs/2015ApJ...809..160F} {809, 160}

\bibitem[\protect\citeauthoryear{{Finzell} et~al.,}{{Finzell}
  et~al.}{2017}]{Finzell+2017}
{Finzell} T.,  et~al., 2017, preprint, \href
  {http://adsabs.harvard.edu/abs/2017arXiv170103094F} {} (\mn@eprint {arXiv}
  {1701.03094})

\bibitem[\protect\citeauthoryear{{Fox} et~al.,}{{Fox} et~al.}{2009}]{Fox09}
{Fox} O.,  et~al., 2009, \mn@doi [\apj] {10.1088/0004-637X/691/1/650}, \href
  {http://adsabs.harvard.edu/abs/2009ApJ...691..650F} {691, 650}

\bibitem[\protect\citeauthoryear{{Friedjung}}{{Friedjung}}{1987}]{Friedjung87}
{Friedjung} M.,  1987, \aap, \href
  {http://adsabs.harvard.edu/abs/1987A%26A...180..155F} {180, 155}

\bibitem[\protect\citeauthoryear{{Friedjung}, {Miko{\l}ajewska}  \&
  {Miko{\l}ajewski}}{{Friedjung} et~al.}{1999}]{Friedjung+99}
{Friedjung} M.,  {Miko{\l}ajewska} J.,   {Miko{\l}ajewski} M.,  1999, \aap,
  \href {http://adsabs.harvard.edu/abs/1999A%26A...348..475F} {348, 475}

\bibitem[\protect\citeauthoryear{{Gehrz}}{{Gehrz}}{2008}]{Gehrz08}
{Gehrz} R.~D.,  2008, {in Classical Novae}

\bibitem[\protect\citeauthoryear{{Gehrz}, {Grasdalen}, {Hackwell}  \&
  {Ney}}{{Gehrz} et~al.}{1980}]{Gehrz+80}
{Gehrz} R.~D.,  {Grasdalen} G.~L.,  {Hackwell} J.~A.,   {Ney} E.~P.,  1980,
  \mn@doi [\apj] {10.1086/157934}, \href
  {http://adsabs.harvard.edu/abs/1980ApJ...237..855G} {237, 855}

\bibitem[\protect\citeauthoryear{{Gehrz}, {Grasdalen}  \& {Hackwell}}{{Gehrz}
  et~al.}{1985}]{Gehrz+85}
{Gehrz} R.~D.,  {Grasdalen} G.~L.,   {Hackwell} J.~A.,  1985, \mn@doi [\apjl]
  {10.1086/184564}, \href {http://adsabs.harvard.edu/abs/1985ApJ...298L..47G}
  {298, L47}

\bibitem[\protect\citeauthoryear{{Gehrz}, {Jones}, {Woodward}, {Greenhouse},
  {Wagner}, {Harrison}, {Hayward}  \& {Benson}}{{Gehrz}
  et~al.}{1992}]{Gehrz+92}
{Gehrz} R.~D.,  {Jones} T.~J.,  {Woodward} C.~E.,  {Greenhouse} M.~A.,
  {Wagner} R.~M.,  {Harrison} T.~E.,  {Hayward} T.~L.,   {Benson} J.,  1992,
  \mn@doi [\apj] {10.1086/172029}, \href
  {http://adsabs.harvard.edu/abs/1992ApJ...400..671G} {400, 671}

\bibitem[\protect\citeauthoryear{{Gehrz}, {Truran}, {Williams}  \&
  {Starrfield}}{{Gehrz} et~al.}{1998}]{Gehrz+98}
{Gehrz} R.~D.,  {Truran} J.~W.,  {Williams} R.~E.,   {Starrfield} S.,  1998,
  \mn@doi [\pasp] {10.1086/316107}, \href
  {http://adsabs.harvard.edu/abs/1998PASP..110....3G} {110, 3}

\bibitem[\protect\citeauthoryear{{Geisel}, {Kleinmann}  \& {Low}}{{Geisel}
  et~al.}{1970}]{Geisel+70}
{Geisel} S.~L.,  {Kleinmann} D.~E.,   {Low} F.~J.,  1970, \mn@doi [\apjl]
  {10.1086/180579}, \href {http://adsabs.harvard.edu/abs/1970ApJ...161L.101G}
  {161, L101}

\bibitem[\protect\citeauthoryear{{Helton}}{{Helton}}{2010a}]{Helton2010}
{Helton} L.~A.,  2010a, PhD thesis, University of Minnesota

\bibitem[\protect\citeauthoryear{{Helton}}{{Helton}}{2010b}]{Helton10}
{Helton} L.~A.,  2010b, PhD thesis, University of Minnesota

\bibitem[\protect\citeauthoryear{{Helton} et~al.,}{{Helton}
  et~al.}{2012}]{Helton+12}
{Helton} L.~A.,  et~al., 2012, \mn@doi [\apj] {10.1088/0004-637X/755/1/37},
  \href {http://adsabs.harvard.edu/abs/2012ApJ...755...37H} {755, 37}

\bibitem[\protect\citeauthoryear{{Hillman}, {Prialnik}, {Kovetz}, {Shara}  \&
  {Neill}}{{Hillman} et~al.}{2014}]{Hillman+14}
{Hillman} Y.,  {Prialnik} D.,  {Kovetz} A.,  {Shara} M.~M.,   {Neill} J.~D.,
  2014, \mn@doi [\mnras] {10.1093/mnras/stt2027}, \href
  {http://adsabs.harvard.edu/abs/2014MNRAS.437.1962H} {437, 1962}

\bibitem[\protect\citeauthoryear{{Johnson}, {Friedlander}  \& {Katz}}{{Johnson}
  et~al.}{1993}]{Johnson93}
{Johnson} D.~J.,  {Friedlander} M.~W.,   {Katz} J.~I.,  1993, \mn@doi [\apj]
  {10.1086/172552}, \href {http://adsabs.harvard.edu/abs/1993ApJ...407..714J}
  {407, 714}

\bibitem[\protect\citeauthoryear{{Kamae}, {Karlsson}, {Mizuno}, {Abe}  \&
  {Koi}}{{Kamae} et~al.}{2006}]{Kamae+06}
{Kamae} T.,  {Karlsson} N.,  {Mizuno} T.,  {Abe} T.,   {Koi} T.,  2006, \mn@doi
  [\apj] {10.1086/505189}, \href
  {http://adsabs.harvard.edu/abs/2006ApJ...647..692K} {647, 692}

\bibitem[\protect\citeauthoryear{{Kato} \& {Hachisu}}{{Kato} \&
  {Hachisu}}{1994}]{Kato&Hachisu94}
{Kato} M.,  {Hachisu} I.,  1994, \mn@doi [\apj] {10.1086/175041}, \href
  {http://adsabs.harvard.edu/abs/1994ApJ...437..802K} {437, 802}

\bibitem[\protect\citeauthoryear{{Keith} \& {Lazzati}}{{Keith} \&
  {Lazzati}}{2011}]{KeithLazzati}
{Keith} A.~C.,  {Lazzati} D.,  2011, \mn@doi [\mnras]
  {10.1111/j.1365-2966.2010.17478.x}, \href
  {http://adsabs.harvard.edu/abs/2011MNRAS.410..685K} {410, 685}

\bibitem[\protect\citeauthoryear{{Kochanek}}{{Kochanek}}{2014}]{Kochanek14}
{Kochanek} C.~S.,  2014, preprint, \href
  {http://adsabs.harvard.edu/abs/2014arXiv1407.7856K} {} (\mn@eprint {arXiv}
  {1407.7856})

\bibitem[\protect\citeauthoryear{{Kozasa} \& {Hasegawa}}{{Kozasa} \&
  {Hasegawa}}{1987}]{KozasaHasegawa87}
{Kozasa} T.,  {Hasegawa} H.,  1987, \mn@doi [Progress of Theoretical Physics]
  {10.1143/PTP.77.1402}, \href
  {http://adsabs.harvard.edu/abs/1987PThPh..77.1402K} {77, 1402}

\bibitem[\protect\citeauthoryear{{Krautter}}{{Krautter}}{2008}]{Krautter08}
{Krautter} J.,  2008, in {Evans} A.,  {Bode} M.~F.,  {O'Brien} T.~J.,
  {Darnley} M.~J.,  eds,  Astronomical Society of the Pacific Conference Series
  Vol. 401, RS Ophiuchi (2006) and the Recurrent Nova Phenomenon. p.~139

\bibitem[\protect\citeauthoryear{{Lazzati} \& {Heger}}{{Lazzati} \&
  {Heger}}{2016}]{Lazzati&Heger16}
{Lazzati} D.,  {Heger} A.,  2016, \mn@doi [\apj] {10.3847/0004-637X/817/2/134},
  \href {http://adsabs.harvard.edu/abs/2016ApJ...817..134L} {817, 134}

\bibitem[\protect\citeauthoryear{{Livio}, {Shankar}, {Burkert}  \&
  {Truran}}{{Livio} et~al.}{1990}]{Livio+90}
{Livio} M.,  {Shankar} A.,  {Burkert} A.,   {Truran} J.~W.,  1990, \mn@doi
  [\apj] {10.1086/168836}, \href
  {http://adsabs.harvard.edu/abs/1990ApJ...356..250L} {356, 250}

\bibitem[\protect\citeauthoryear{{Lloyd}, {O'Brien}  \& {Bode}}{{Lloyd}
  et~al.}{1997}]{Lloyd+97}
{Lloyd} H.~M.,  {O'Brien} T.~J.,   {Bode} M.~F.,  1997, \mnras, \href
  {http://adsabs.harvard.edu/abs/1997MNRAS.284..137L} {284, 137}

\bibitem[\protect\citeauthoryear{{Lynch} et~al.}{{Lynch}
  et~al.}{2008}]{Lynch+08}
{Lynch} D.~K.,  et~al., 2008, \mn@doi [\aj] {10.1088/0004-6256/136/5/1815},
  \href {http://adsabs.harvard.edu/abs/2008AJ....136.1815L} {136, 1815}

\bibitem[\protect\citeauthoryear{{Mason}, {Gehrz}, {Woodward}, {Smilowitz},
  {Hayward}  \& {Houck}}{{Mason} et~al.}{1998}]{Mason+98}
{Mason} C.~G.,  {Gehrz} R.~D.,  {Woodward} C.~E.,  {Smilowitz} J.~B.,
  {Hayward} T.~L.,   {Houck} J.~R.,  1998, \mn@doi [\apj] {10.1086/305220},
  \href {http://adsabs.harvard.edu/abs/1998ApJ...494..783M} {494, 783}

\bibitem[\protect\citeauthoryear{{Mauney}, {Buongiorno Nardelli}  \&
  {Lazzati}}{{Mauney} et~al.}{2015}]{Mauney+2015}
{Mauney} C.,  {Buongiorno Nardelli} M.,   {Lazzati} D.,  2015, \mn@doi [\apj]
  {10.1088/0004-637X/800/1/30}, \href
  {http://adsabs.harvard.edu/abs/2015ApJ...800...30M} {800, 30}

\bibitem[\protect\citeauthoryear{{McLaughlin}}{{McLaughlin}}{1935}]{McLaughlin35}
{McLaughlin} D.~B.,  1935, Popular Astronomy, \href
  {http://adsabs.harvard.edu/abs/1935PA.....43..265M} {43, 265}

\bibitem[\protect\citeauthoryear{{Metzger}, {Hasco{\"e}t}, {Vurm},
  {Beloborodov}, {Chomiuk}, {Sokoloski}  \& {Nelson}}{{Metzger}
  et~al.}{2014}]{Metzger+14}
{Metzger} B.~D.,  {Hasco{\"e}t} R.,  {Vurm} I.,  {Beloborodov} A.~M.,
  {Chomiuk} L.,  {Sokoloski} J.~L.,   {Nelson} T.,  2014, \mn@doi [\mnras]
  {10.1093/mnras/stu844}, \href
  {http://adsabs.harvard.edu/abs/2014MNRAS.442..713M} {442, 713}

\bibitem[\protect\citeauthoryear{{Metzger}, {Finzell}, {Vurm}, {Hasco{\"e}t},
  {Beloborodov}  \& {Chomiuk}}{{Metzger} et~al.}{2015a}]{Metzger2015}
{Metzger} B.~D.,  {Finzell} T.,  {Vurm} I.,  {Hasco{\"e}t} R.,  {Beloborodov}
  A.~M.,   {Chomiuk} L.,  2015a, \mn@doi [\mnras] {10.1093/mnras/stv742}, \href
  {http://adsabs.harvard.edu/abs/2015MNRAS.450.2739M} {450, 2739}

\bibitem[\protect\citeauthoryear{{Metzger}, {Finzell}, {Vurm}, {Hasco{\"e}t},
  {Beloborodov}  \& {Chomiuk}}{{Metzger} et~al.}{2015b}]{Metzger+15}
{Metzger} B.~D.,  {Finzell} T.,  {Vurm} I.,  {Hasco{\"e}t} R.,  {Beloborodov}
  A.~M.,   {Chomiuk} L.,  2015b, \mn@doi [\mnras] {10.1093/mnras/stv742}, \href
  {http://adsabs.harvard.edu/abs/2015MNRAS.450.2739M} {450, 2739}

\bibitem[\protect\citeauthoryear{{Metzger}, {Caprioli}, {Vurm}, {Beloborodov},
  {Bartos}  \& {Vlasov}}{{Metzger} et~al.}{2016}]{Metzger+16}
{Metzger} B.~D.,  {Caprioli} D.,  {Vurm} I.,  {Beloborodov} A.~M.,  {Bartos}
  I.,   {Vlasov} A.,  2016, \mn@doi [\mnras] {10.1093/mnras/stw123}, \href
  {http://adsabs.harvard.edu/abs/2016MNRAS.457.1786M} {457, 1786}

\bibitem[\protect\citeauthoryear{{Morisset} \& {Pequignot}}{{Morisset} \&
  {Pequignot}}{1996}]{Morisset&Pequignot96}
{Morisset} C.,  {Pequignot} D.,  1996, \aap, \href
  {http://adsabs.harvard.edu/abs/1996A%26A...312..135M} {312, 135}

\bibitem[\protect\citeauthoryear{{Mukai} \& {Ishida}}{{Mukai} \&
  {Ishida}}{2001}]{Mukai&Ishida01}
{Mukai} K.,  {Ishida} M.,  2001, \mn@doi [\apj] {10.1086/320220}, \href
  {http://adsabs.harvard.edu/abs/2001ApJ...551.1024M} {551, 1024}

\bibitem[\protect\citeauthoryear{{Mukai}, {Orio}  \& {Della Valle}}{{Mukai}
  et~al.}{2008}]{Mukai+08}
{Mukai} K.,  {Orio} M.,   {Della Valle} M.,  2008, \mn@doi [\apj]
  {10.1086/529362}, \href {http://adsabs.harvard.edu/abs/2008ApJ...677.1248M}
  {677, 1248}

\bibitem[\protect\citeauthoryear{{Munari}, {Siviero}, {Henden}, {Ochner},
  {Simoncelli}, {Tomasoni}, {Moschini}  \& {Dallaporta}}{{Munari}
  et~al.}{2007}]{Munari+07}
{Munari} U.,  {Siviero} A.,  {Henden} A.,  {Ochner} P.,  {Simoncelli} C.,
  {Tomasoni} S.,  {Moschini} F.,   {Dallaporta} S.,  2007, Central Bureau
  Electronic Telegrams, \href
  {http://adsabs.harvard.edu/abs/2007CBET.1099....1M} {1099}

\bibitem[\protect\citeauthoryear{{Nagashima} et~al.,}{{Nagashima}
  et~al.}{2013}]{Nagashima+2013}
{Nagashima} M.,  et~al., 2013, in {Di Stefano} R.,  {Orio} M.,   {Moe} M.,
  eds,  IAU Symposium Vol. 281, Binary Paths to Type Ia Supernovae Explosions.
  pp 121--123, \mn@doi{10.1017/S1743921312014810}

\bibitem[\protect\citeauthoryear{{Nahar}}{{Nahar}}{1996}]{Nahar96}
{Nahar} S.~N.,  1996, \mn@doi [\apjs] {10.1086/192336}, \href
  {http://adsabs.harvard.edu/abs/1996ApJS..106..213N} {106, 213}

\bibitem[\protect\citeauthoryear{{Ney} \& {Hatfield}}{{Ney} \&
  {Hatfield}}{1978}]{Ney&Hatfield78}
{Ney} E.~P.,  {Hatfield} B.~F.,  1978, \mn@doi [\apjl] {10.1086/182618}, \href
  {http://adsabs.harvard.edu/abs/1978ApJ...219L.111N} {219, L111}

\bibitem[\protect\citeauthoryear{{Nofar}, {Shaviv}  \& {Starrfield}}{{Nofar}
  et~al.}{1991}]{Nofar+91}
{Nofar} I.,  {Shaviv} G.,   {Starrfield} S.,  1991, \mn@doi [\apj]
  {10.1086/169772}, \href {http://adsabs.harvard.edu/abs/1991ApJ...369..440N}
  {369, 440}

\bibitem[\protect\citeauthoryear{{Nozawa}, {Kozasa}, {Umeda}, {Maeda}  \&
  {Nomoto}}{{Nozawa} et~al.}{2003}]{Nozawa2003}
{Nozawa} T.,  {Kozasa} T.,  {Umeda} H.,  {Maeda} K.,   {Nomoto} K.,  2003,
  \mn@doi [\apj] {10.1086/379011}, \href
  {http://adsabs.harvard.edu/abs/2003ApJ...598..785N} {598, 785}

\bibitem[\protect\citeauthoryear{{Orlando} \& {Drake}}{{Orlando} \&
  {Drake}}{2012}]{Orlando&Drake12}
{Orlando} S.,  {Drake} J.~J.,  2012, \mn@doi [\mnras]
  {10.1111/j.1365-2966.2011.19880.x}, \href
  {http://adsabs.harvard.edu/abs/2012MNRAS.419.2329O} {419, 2329}

\bibitem[\protect\citeauthoryear{{Orlando}, {Drake}  \& {Laming}}{{Orlando}
  et~al.}{2009}]{Orlando+09}
{Orlando} S.,  {Drake} J.~J.,   {Laming} J.~M.,  2009, \mn@doi [\aap]
  {10.1051/0004-6361:200810109}, \href
  {http://adsabs.harvard.edu/abs/2009A%26A...493.1049O} {493, 1049}

\bibitem[\protect\citeauthoryear{{Osterbrock} \& {Ferland}}{{Osterbrock} \&
  {Ferland}}{2006}]{Osterbrock&Ferland06}
{Osterbrock} D.~E.,  {Ferland} G.~J.,  2006, {Astrophysics of gaseous nebulae
  and active galactic nuclei}

\bibitem[\protect\citeauthoryear{{Pontefract} \& {Rawlings}}{{Pontefract} \&
  {Rawlings}}{2004}]{Pontefract&Rawlings04}
{Pontefract} M.,  {Rawlings} J.~M.~C.,  2004, \mn@doi [\mnras]
  {10.1111/j.1365-2966.2004.07330.x}, \href
  {http://adsabs.harvard.edu/abs/2004MNRAS.347.1294P} {347, 1294}

\bibitem[\protect\citeauthoryear{{Rawlings}}{{Rawlings}}{1988}]{Rawlings88}
{Rawlings} J.~M.~C.,  1988, \mn@doi [\mnras] {10.1093/mnras/232.3.507}, \href
  {http://adsabs.harvard.edu/abs/1988MNRAS.232..507R} {232, 507}

\bibitem[\protect\citeauthoryear{{Rawlings} \& {Evans}}{{Rawlings} \&
  {Evans}}{2008}]{Rawlings&Evans08}
{Rawlings} J.,  {Evans} A.,  2008, {in Classical Novae}

\bibitem[\protect\citeauthoryear{{Rawlings} \& {Williams}}{{Rawlings} \&
  {Williams}}{1989}]{Rawlings&Williams89}
{Rawlings} J.~M.~C.,  {Williams} D.~A.,  1989, \mn@doi [\mnras]
  {10.1093/mnras/240.3.729}, \href
  {http://adsabs.harvard.edu/abs/1989MNRAS.240..729R} {240, 729}

\bibitem[\protect\citeauthoryear{{Ribeiro}, {Munari}  \& {Valisa}}{{Ribeiro}
  et~al.}{2013}]{Ribeiro+13}
{Ribeiro} V.~A.~R.~M.,  {Munari} U.,   {Valisa} P.,  2013, \mn@doi [\apj]
  {10.1088/0004-637X/768/1/49}, \href
  {http://adsabs.harvard.edu/abs/2013ApJ...768...49R} {768, 49}

\bibitem[\protect\citeauthoryear{{Roche}, {Aitken}  \& {Whitmore}}{{Roche}
  et~al.}{1984}]{Roche+84}
{Roche} P.~F.,  {Aitken} D.~K.,   {Whitmore} B.,  1984, \mn@doi [\mnras]
  {10.1093/mnras/211.3.535}, \href
  {http://adsabs.harvard.edu/abs/1984MNRAS.211..535R} {211, 535}

\bibitem[\protect\citeauthoryear{{Roth} \& {Garcia-Rosales}}{{Roth} \&
  {Garcia-Rosales}}{1996}]{Roth&GarciaRosales96}
{Roth} J.,  {Garcia-Rosales} C.,  1996, \mn@doi [Nuclear Fusion]
  {10.1088/0029-5515/36/12/I05}, \href
  {http://adsabs.harvard.edu/abs/1996NucFu..36.1647R} {36, 1647}

\bibitem[\protect\citeauthoryear{{Rudy}, {Dimpfl}, {Lynch}, {Mazuk},
  {Venturini}, {Wilson}, {Puetter}  \& {Perry}}{{Rudy} et~al.}{2003}]{Rudy+03}
{Rudy} R.~J.,  {Dimpfl} W.~L.,  {Lynch} D.~K.,  {Mazuk} S.,  {Venturini} C.~C.,
   {Wilson} J.~C.,  {Puetter} R.~C.,   {Perry} R.~B.,  2003, \mn@doi [\apj]
  {10.1086/378083}, \href {http://adsabs.harvard.edu/abs/2003ApJ...596.1229R}
  {596, 1229}

\bibitem[\protect\citeauthoryear{{Saizar} \& {Ferland}}{{Saizar} \&
  {Ferland}}{1994}]{Saizar&Ferland94}
{Saizar} P.,  {Ferland} G.~J.,  1994, \mn@doi [\apj] {10.1086/174019}, \href
  {http://adsabs.harvard.edu/abs/1994ApJ...425..755S} {425, 755}

\bibitem[\protect\citeauthoryear{{Sakon} et~al.,}{{Sakon}
  et~al.}{2016}]{Sakon+16}
{Sakon} I.,  et~al., 2016, \mn@doi [\apj] {10.3847/0004-637X/817/2/145}, \href
  {http://adsabs.harvard.edu/abs/2016ApJ...817..145S} {817, 145}

\bibitem[\protect\citeauthoryear{{Schaefer} et~al.}{{Schaefer}
  et~al.}{2014}]{Schaefer+14}
{Schaefer} G.~H.,  et~al., 2014, \mn@doi [\nat] {10.1038/nature13834}, \href
  {http://adsabs.harvard.edu/abs/2014Natur.515..234S} {515, 234}

\bibitem[\protect\citeauthoryear{{Schure}, {Kosenko}, {Kaastra}, {Keppens}  \&
  {Vink}}{{Schure} et~al.}{2009}]{Schure2009}
{Schure} K.~M.,  {Kosenko} D.,  {Kaastra} J.~S.,  {Keppens} R.,   {Vink} J.,
  2009, \mn@doi [\aap] {10.1051/0004-6361/200912495}, \href
  {http://adsabs.harvard.edu/abs/2009A%26A...508..751S} {508, 751}

\bibitem[\protect\citeauthoryear{{Schwarz}}{{Schwarz}}{2002a}]{Schwarz02}
{Schwarz} G.~J.,  2002a, \mn@doi [\apj] {10.1086/342234}, \href
  {http://adsabs.harvard.edu/abs/2002ApJ...577..940S} {577, 940}

\bibitem[\protect\citeauthoryear{{Schwarz}}{{Schwarz}}{2002b}]{Schwarz+02}
{Schwarz} G.~J.,  2002b, \mn@doi [\apj] {10.1086/342234}, \href
  {http://adsabs.harvard.edu/abs/2002ApJ...577..940S} {577, 940}

\bibitem[\protect\citeauthoryear{{Schwarz}, {Hauschildt}, {Starrfield},
  {Baron}, {Allard}, {Shore}  \& {Sonneborn}}{{Schwarz}
  et~al.}{1997}]{Schwarz+97}
{Schwarz} G.~J.,  {Hauschildt} P.~H.,  {Starrfield} S.,  {Baron} E.,  {Allard}
  F.,  {Shore} S.~N.,   {Sonneborn} G.,  1997, \mn@doi [\mnras]
  {10.1093/mnras/284.3.669}, \href
  {http://adsabs.harvard.edu/abs/1997MNRAS.284..669S} {284, 669}

\bibitem[\protect\citeauthoryear{{Schwarz}, {Shore}, {Starrfield}  \&
  {Vanlandingham}}{{Schwarz} et~al.}{2007}]{Schwarz+07}
{Schwarz} G.~J.,  {Shore} S.~N.,  {Starrfield} S.,   {Vanlandingham} K.~M.,
  2007, \mn@doi [\apj] {10.1086/510661}, \href
  {http://adsabs.harvard.edu/abs/2007ApJ...657..453S} {657, 453}

\bibitem[\protect\citeauthoryear{{Schwarz} et~al.,}{{Schwarz}
  et~al.}{2011}]{Schwarz+11}
{Schwarz} G.~J.,  et~al., 2011, \mn@doi [\apjs] {10.1088/0067-0049/197/2/31},
  \href {http://adsabs.harvard.edu/abs/2011ApJS..197...31S} {197, 31}

\bibitem[\protect\citeauthoryear{{Scott}, {Evans}  \& {Rawlings}}{{Scott}
  et~al.}{1994}]{Scott+94}
{Scott} A.~D.,  {Evans} A.,   {Rawlings} J.~M.~C.,  1994, \mn@doi [\mnras]
  {10.1093/mnras/269.1.21L}, \href
  {http://adsabs.harvard.edu/abs/1994MNRAS.269L..21S} {269, L21}

\bibitem[\protect\citeauthoryear{{Seaquist}, {Duric}, {Israel}, {Spoelstra},
  {Ulich}  \& {Gregory}}{{Seaquist} et~al.}{1980}]{Seaquist+80}
{Seaquist} E.~R.,  {Duric} N.,  {Israel} F.~P.,  {Spoelstra} T.~A.~T.,  {Ulich}
  B.~L.,   {Gregory} P.~C.,  1980, \mn@doi [\aj] {10.1086/112672}, \href
  {http://adsabs.harvard.edu/abs/1980AJ.....85..283S} {85, 283}

\bibitem[\protect\citeauthoryear{{Seaton}, {Yan}, {Mihalas}  \&
  {Pradhan}}{{Seaton} et~al.}{1994}]{Seaton+94}
{Seaton} M.~J.,  {Yan} Y.,  {Mihalas} D.,   {Pradhan} A.~K.,  1994, \mn@doi
  [\mnras] {10.1093/mnras/266.4.805}, \href
  {http://adsabs.harvard.edu/abs/1994MNRAS.266..805S} {266, 805}

\bibitem[\protect\citeauthoryear{{Semenov}, {Henning}, {Helling}, {Ilgner}  \&
  {Sedlmayr}}{{Semenov} et~al.}{2003}]{Semenov+03}
{Semenov} D.,  {Henning} T.,  {Helling} C.,  {Ilgner} M.,   {Sedlmayr} E.,
  2003, \mn@doi [\aap] {10.1051/0004-6361:20031279}, \href
  {http://adsabs.harvard.edu/abs/2003A%26A...410..611S} {410, 611}

\bibitem[\protect\citeauthoryear{{Shore}}{{Shore}}{2012}]{Shore12}
{Shore} S.~N.,  2012, Bulletin of the Astronomical Society of India, \href
  {http://adsabs.harvard.edu/abs/2012BASI...40..185S} {40, 185}

\bibitem[\protect\citeauthoryear{{Shore}, {Starrfield}, {Gonzalez-Riestrat},
  {Hauschildt}  \& {Sonneborn}}{{Shore} et~al.}{1994}]{Shore+94}
{Shore} S.~N.,  {Starrfield} S.,  {Gonzalez-Riestrat} R.,  {Hauschildt} P.~H.,
   {Sonneborn} G.,  1994, \mn@doi [\nat] {10.1038/369539a0}, \href
  {http://adsabs.harvard.edu/abs/1994Natur.369..539S} {369, 539}

\bibitem[\protect\citeauthoryear{{Shore}, {De Gennaro Aquino}, {Schwarz},
  {Augusteijn}, {Cheung}, {Walter}  \& {Starrfield}}{{Shore}
  et~al.}{2013}]{Shore+13}
{Shore} S.~N.,  {De Gennaro Aquino} I.,  {Schwarz} G.~J.,  {Augusteijn} T.,
  {Cheung} C.~C.,  {Walter} F.~M.,   {Starrfield} S.,  2013, \mn@doi [\aap]
  {10.1051/0004-6361/201321095}, \href
  {http://adsabs.harvard.edu/abs/2013A%26A...553A.123S} {553, A123}

\bibitem[\protect\citeauthoryear{{Sirko} \& {Goodman}}{{Sirko} \&
  {Goodman}}{2003}]{Sirko&Goodman03}
{Sirko} E.,  {Goodman} J.,  2003, \mn@doi [\mnras]
  {10.1046/j.1365-8711.2003.06431.x}, \href
  {http://adsabs.harvard.edu/abs/2003MNRAS.341..501S} {341, 501}

\bibitem[\protect\citeauthoryear{{Smith}}{{Smith}}{2010}]{Smith10}
{Smith} N.,  2010, \mn@doi [\mnras] {10.1111/j.1365-2966.2009.15901.x}, \href
  {http://adsabs.harvard.edu/abs/2010MNRAS.402..145S} {402, 145}

\bibitem[\protect\citeauthoryear{{Smith}, {Aitken}, {Roche}  \&
  {Wright}}{{Smith} et~al.}{1995}]{Smith+95}
{Smith} C.~H.,  {Aitken} D.~K.,  {Roche} P.~F.,   {Wright} C.~M.,  1995,
  \mn@doi [\mnras] {10.1093/mnras/277.1.259}, \href
  {http://adsabs.harvard.edu/abs/1995MNRAS.277..259S} {277, 259}

\bibitem[\protect\citeauthoryear{{Smith}, {Foley}  \& {Filippenko}}{{Smith}
  et~al.}{2008}]{Smith+08}
{Smith} N.,  {Foley} R.~J.,   {Filippenko} A.~V.,  2008, \mn@doi [\apj]
  {10.1086/587860}, \href {http://adsabs.harvard.edu/abs/2008ApJ...680..568S}
  {680, 568}

\bibitem[\protect\citeauthoryear{{Snijders}, {Batt}, {Roche}, {Seaton},
  {Morton}, {Spoelstra}  \& {Blades}}{{Snijders} et~al.}{1987}]{Snijders+87}
{Snijders} M.~A.~J.,  {Batt} T.~J.,  {Roche} P.~F.,  {Seaton} M.~J.,  {Morton}
  D.~C.,  {Spoelstra} T.~A.~T.,   {Blades} J.~C.,  1987, \mn@doi [\mnras]
  {10.1093/mnras/228.2.329}, \href
  {http://adsabs.harvard.edu/abs/1987MNRAS.228..329S} {228, 329}

\bibitem[\protect\citeauthoryear{{Sokoloski}, {Rupen}  \&
  {Mioduszewski}}{{Sokoloski} et~al.}{2008}]{Sokoloski+08}
{Sokoloski} J.~L.,  {Rupen} M.~P.,   {Mioduszewski} A.~J.,  2008, \mn@doi
  [\apjl] {10.1086/592602}, \href
  {http://adsabs.harvard.edu/abs/2008ApJ...685L.137S} {685, L137}

\bibitem[\protect\citeauthoryear{{Starrfield}, {Truran}, {Wiescher}  \&
  {Sparks}}{{Starrfield} et~al.}{1998}]{Starrfield+98}
{Starrfield} S.,  {Truran} J.~W.,  {Wiescher} M.~C.,   {Sparks} W.~M.,  1998,
  \mn@doi [\mnras] {10.1046/j.1365-8711.1998.01312.x}, \href
  {http://adsabs.harvard.edu/abs/1998MNRAS.296..502S} {296, 502}

\bibitem[\protect\citeauthoryear{{Todini} \& {Ferrara}}{{Todini} \&
  {Ferrara}}{2001}]{Todini&Ferrara01}
{Todini} P.,  {Ferrara} A.,  2001, \mn@doi [\mnras]
  {10.1046/j.1365-8711.2001.04486.x}, \href
  {http://adsabs.harvard.edu/abs/2001MNRAS.325..726T} {325, 726}

\bibitem[\protect\citeauthoryear{{Tomida}, {Tomisaka}, {Matsumoto}, {Hori},
  {Okuzumi}, {Machida}  \& {Saigo}}{{Tomida} et~al.}{2013}]{Tomida+13}
{Tomida} K.,  {Tomisaka} K.,  {Matsumoto} T.,  {Hori} Y.,  {Okuzumi} S.,
  {Machida} M.~N.,   {Saigo} K.,  2013, \mn@doi [\apj]
  {10.1088/0004-637X/763/1/6}, \href
  {http://adsabs.harvard.edu/abs/2013ApJ...763....6T} {763, 6}

\bibitem[\protect\citeauthoryear{{Tomida}, {Okuzumi}  \& {Machida}}{{Tomida}
  et~al.}{2015}]{Tomida+15}
{Tomida} K.,  {Okuzumi} S.,   {Machida} M.~N.,  2015, \mn@doi [\apj]
  {10.1088/0004-637X/801/2/117}, \href
  {http://adsabs.harvard.edu/abs/2015ApJ...801..117T} {801, 117}

\bibitem[\protect\citeauthoryear{{Toraskar}, {Mac Low}, {Shara}  \&
  {Zurek}}{{Toraskar} et~al.}{2013}]{Toraskar13}
{Toraskar} J.,  {Mac Low} M.-M.,  {Shara} M.~M.,   {Zurek} D.~R.,  2013,
  \mn@doi [\apj] {10.1088/0004-637X/768/1/48}, \href
  {http://adsabs.harvard.edu/abs/2013ApJ...768...48T} {768, 48}

\bibitem[\protect\citeauthoryear{{Vishniac}}{{Vishniac}}{1983}]{Vishniac83}
{Vishniac} E.~T.,  1983, \mn@doi [\apj] {10.1086/161433}, \href
  {http://adsabs.harvard.edu/abs/1983ApJ...274..152V} {274, 152}

\bibitem[\protect\citeauthoryear{{Vlasov}, {Vurm}  \& {Metzger}}{{Vlasov}
  et~al.}{2016}]{Vlasov+2016}
{Vlasov} A.,  {Vurm} I.,   {Metzger} B.~D.,  2016, \mn@doi [\mnras]
  {10.1093/mnras/stw1949}, \href
  {http://adsabs.harvard.edu/abs/2016MNRAS.463..394V} {463, 394}

\bibitem[\protect\citeauthoryear{{Vurm} \& {Metzger}}{{Vurm} \&
  {Metzger}}{2016}]{VurmMetzger2016}
{Vurm} I.,  {Metzger} B.~D.,  2016, preprint, \href
  {http://adsabs.harvard.edu/abs/2016arXiv161104532V} {} (\mn@eprint {arXiv}
  {1611.04532})

\bibitem[\protect\citeauthoryear{{Walter}, {Battisti}, {Towers}, {Bond}  \&
  {Stringfellow}}{{Walter} et~al.}{2012}]{Walter+12}
{Walter} F.~M.,  {Battisti} A.,  {Towers} S.~E.,  {Bond} H.~E.,
  {Stringfellow} G.~S.,  2012, \mn@doi [\pasp] {10.1086/668404}, \href
  {http://adsabs.harvard.edu/abs/2012PASP..124.1057W} {124, 1057}

\bibitem[\protect\citeauthoryear{{Waters}}{{Waters}}{2004}]{Waters04}
{Waters} L.~B.~F.~M.,  2004, in {Witt} A.~N.,  {Clayton} G.~C.,   {Draine}
  B.~T.,  eds,  Astronomical Society of the Pacific Conference Series Vol. 309,
  Astrophysics of Dust. p.~229

\bibitem[\protect\citeauthoryear{{Weston} et~al.,}{{Weston}
  et~al.}{2015}]{Weston+15}
{Weston} J.~H.~S.,  et~al., 2015, preprint, \href
  {http://adsabs.harvard.edu/abs/2015arXiv150505879W} {} (\mn@eprint {arXiv}
  {1505.05879})

\bibitem[\protect\citeauthoryear{{Williams}}{{Williams}}{1992}]{Williams92}
{Williams} R.~E.,  1992, \mn@doi [\apj] {10.1086/171409}, \href
  {http://adsabs.harvard.edu/abs/1992ApJ...392...99W} {392, 99}

\bibitem[\protect\citeauthoryear{{Williams} \& {Mason}}{{Williams} \&
  {Mason}}{2010}]{Williams&Mason10}
{Williams} R.,  {Mason} E.,  2010, \mn@doi [\apss] {10.1007/s10509-010-0318-x},
  \href {http://adsabs.harvard.edu/abs/2010Ap%26SS.327..207W} {327, 207}

\bibitem[\protect\citeauthoryear{{Wolf}, {Bildsten}, {Brooks}  \&
  {Paxton}}{{Wolf} et~al.}{2013}]{Wolf+13}
{Wolf} W.~M.,  {Bildsten} L.,  {Brooks} J.,   {Paxton} B.,  2013, \mn@doi
  [\apj] {10.1088/0004-637X/777/2/136}, \href
  {http://adsabs.harvard.edu/abs/2013ApJ...777..136W} {777, 136}

\makeatother
\end{thebibliography}

\end{document}